%% file: sm1.tex
\documentstyle[ltwol,epsfig,axodraw]{article}

\def\NPB{{\em Nucl. Phys.} B }
\def\PLB{{\em Phys. Lett.}  B }
\def\PRL{{\em Phys. Rev. Lett.} }
\def\PRD{{\em Phys. Rev.} D }
\def\ZPC{{\em Z. Phys.} C }

\def\ra{\rightarrow}

\def\al{\alpha}

\def\be{\begin{equation}}
\def\ee{\end{equation}}
\def\bea{\begin{eqnarray}}
\def\eea{\end{eqnarray}}
\newcommand{\beq}{\begin{equation}}
\newcommand{\eeq}{\end{equation}}
\newcommand{\barr}{\begin{array}}
\newcommand{\earr}{\end{array}}
\newcommand{\bc}{\begin{center}}
\newcommand{\ec}{\end{center}}
\newcommand{\btab}{\begin{tabular}}
\newcommand{\etab}{\end{tabular}}
\newcommand{\gv}{\mbox{GeV}}

\newcommand{\nn}{\nonumber}

\newcommand{\dro}{\Delta\rho}
\newcommand{\drqcd}{\delta\rho_{\rm QCD}}
\newcommand{\roro}{\rho^{(2)}}

\newcommand{\alpi}{\frac{\alpha}{\pi}}

\newcommand{\m}{\mu}
\newcommand{\g}{\gamma}
\newcommand{\G}{\Gamma}
\newcommand{\Gmu}{G_{\mu}}
\newcommand{\amu}{a_{\mu}}

\newcommand{\ganu}{\gamma_{\nu}}

\newcommand{\gafi}{\gamma_5}

\newcommand{\Pigr}{\hat{\Pi}^{\gamma}}

\newcommand{\noi}{\noindent}
\newcommand{\epmf}{e^+e^- \rightarrow f\bar{f}}

\newcommand{\epm}{e^+e^-}

\newcommand{\sm}{standard model }
\newcommand{\su}{SU(2)$\times$U(1) }
\newcommand{\siw}{\sin^2\theta_W}

\newcommand{\dal}{\Delta\alpha}

\newcommand{\mz}{M_Z^2}
\newcommand{\mw}{M_W^2}

\newcommand{\real}{\mbox{Re}}

\newcommand{\Dr}{\Delta r}

\newcommand{\veps}{\varepsilon}

\newcommand{\aspi}{\frac{\alpha_s}{\pi}}

\newcommand{\alr}{A_{\rm LR}}
\newcommand{\afb}{A_{\rm FB}}

\newcommand{\ass}{asymmetries }

\newcommand{\sell}{s_{\ell}^2}

\newcommand{\ms}{\overline{MS}}

\newcommand{\nf}{{\mathrm{nf}}}
\def\mathswitchr#1{\relax\ifmmode{\mathrm{#1}}\else$\mathrm{#1}$\fi}
\def\mathswitch#1{\relax\ifmmode#1\else$#1$\fi}

\newcommand{\PW}{\mathswitchr W}

\newcommand{\Pep}{\mathswitchr {e^+}}
\newcommand{\Pem}{\mathswitchr {e^-}}

\begin{document}

\input{titlepage}

\title{STANDARD MODEL THEORY}

\author{W. HOLLIK}

\address{Theoretical Physics Division, 
         CERN, CH-1211 Geneva 23, Switzerland \\
            and \\
         Institut f\"ur Theoretische Physik, Universit\"at Karlsruhe,
         D-76128 Karlsruhe, Germany \\
         E-mail: Wolfgang.Hollik@physik.uni-karlsruhe.de}   


\twocolumn[\maketitle\abstracts{
In this conference report a summary is given on the theoretical 
work that has contributed to provide accurate 
theoretical predictions for testing the \sm in present and future 
experiments. Precision calculations for the  vector boson masses,
for the $Z$ resonance, $W$ pair production, and 
for the $g-2$ of the muon
are reviewed and the theoretical situation for the Higgs sector
is summarized.
 The status of the \sm is discussed 
in the light of the recent 
high and low energy data.
New Physics beyond the \sm is briefly addressed 
as well, with special
emphasis on the minimal supersymmetric standard model.
}]
 
\section{Introduction}

The $\epm$ colliders LEP and the SLC, in operation since summer
1989, have collected an enormous amount of electroweak precision data
on  $Z$ and $W$ bosons \cite{lep,karlen}. The 
$W$ boson properties have in parallel been
determined at the $p \bar{p}$
collider Tevatron with a constant increase 
in accuracy~\cite{wmass,karlen};
after the 
discovery of the top quark there \cite{top},
its mass has been measured  \cite{partridge}
with a precision of better than 3\%, to
$173.8\pm 5.0$ GeV.
The ongoing experiments at LEP 2 and the 
near-future Tevatron upgrade will also, in the coming years,
support us with further increases in precision, in particular on
the mass of the $W$ and the top, and the SLC might continue to improve
the  impressive accuracy already obtained in the electroweak mixing angle.
This stimulating experimental program together with the theoretical 
activities to provide accurate predictions from the standard model
have formed the era of electroweak
precison tests and will keep it alive also in the next years.

\medskip 
The standard
theory of the electroweak interaction 
 is a gauge-invariant quantum field theory with
the symmetry group \su spontaneously broken by the Higgs mechanism.
The possibility to perform perturbative calculations for observable
quantities in terms of a few input parameters is substantially based on
the renormalizability of this class of theories \cite{thooft}.
A certain set of input parameters
has to be taken from experiment.
In the electroweak standard model essentially 
three free parameters are required to describe the gauge bosons
$\g,W^{\pm},Z$,
and their interactions with the fermions.
For a
comparison between theory and experiment,
hence, three independent experimental
input data are required. The most natural choice consists of
the electromagnetic fine structure constant $\al$, the
muon decay constant (Fermi constant) $\Gmu$, and the mass of the $Z$
boson, which has meanwhile been measured with the same accuracy
as the Fermi constant 
\cite{lep,karlen}.
              Other measurable quantities
are predicted in terms of the input data. Each additional precision
experiment, which allows the detection of small deviations from the
lowest-order predictions, can be considered a test of the electroweak
theory at the quantum level.
In the Feynman graph expansion of the scattering amplitude
for a given process the  higher-order terms
show up as diagrams containing closed loops.
 The renormalizability of the \sm  ensures that
it retains its predictive power also  in higher orders.
The higher-order terms
are the quantum effects of the electroweak theory. They  
are  complicated in their concrete form, but
they are finally the consequence of the basic Lagrangian with a
simple structure.
The quantum corrections (or ``radiative corrections'') 
 contain the self-coupling of the vector bosons as well as their
interactions with the Higgs field and the top quark,
and  provide the theoretical basis for electroweak precision
tests. Assuming the validity of the standard model, the presence of the
top quark and the Higgs boson in the loop contributions to electroweak
observables allows an  indirect probe of their mass 
ranges  from comparison with precision data.

\medskip
The generation of high-precision experiments hence imposes
stringent tests on the standard model.
A primordial step strengthening our confidence in the \sm 
has been the discovery 
of the top quark at the Tevatron \cite{top}, at a mass
that agrees with the mass range obtained  indirectly, through
the radiative corrections.
Moreover, with the top mass as an additional precise experimental
data point one can now fully
exploit the virtual sensitivity to the Higgs mass.

\medskip
The experimental
sensitivity in the electroweak observables, at the
level of the quantum effects,  requires the highest standards
on the theoretical side as well. A sizeable amount of work has
contributed, over the recent years, to a steadily rising
improvement of the standard model predictions,
pinning down the theoretical uncertainties to the level 
required for the current interpretation of the precision data.
The availability of both 
highly accurate measurements and theoretical predictions, at the level
of 0.1\% precision and better, provides tests of 
the quantum structure of the standard model, thereby
probing its  still untested scalar sector, 
and simultaneously accesses
alternative scenarios such as the supersymmetric extension 
of the standard model.

\medskip
 The lack
of direct signals from new physics beyond the \sm 
makes the high-precision experiments
a unique tool also in the search for indirect effects:
through possible deviations of the experimental results
from the theoretical predictions of the minimal standard model.
Since such deviations are expected to be small,
it is decisive  to have 
 the standard loop effects in the precision observables under control.

\bigskip
This review contains a discussion of the
theoretical developments for testing the electroweak theory,
the status of the \sm in view of the most recent high and low
energy data, and the implications for the 
Higgs boson. New Physics is briefly included, with main emphasis 
on the minimal supersymmetric standard model.

\section{Status of precision calculations}
\subsection{Basic ingredients in radiative corrections}

The possibility of performing precision tests is based
on the formulation of the \sm as a renormalizable quantum field
theory preserving its predictive power beyond tree-level
calculations. With the experimental accuracy 
being sensitive to the loop-induced quantum effects, 
also the Higgs sector of the \sm
is being probed. The higher-order terms
induce the sensitivity of electroweak observables
to the top and Higgs mass $m_t, M_H$
and to the strong coupling constant $\al_s$.

\smallskip 
The calculation of electroweak observables in higher orders
requires the concept of renormalization to get rid of the
divergences in the Feynman integral evaluation and to define 
the physical input parameters.  
In QED and in the electroweak theory
the classical Thomson scattering and the particle masses
set natural scales where the parameters 
 $e=\sqrt{4\pi\alpha}$  and the electron, muon, \dots masses
can be defined.
In the electroweak standard model a distinguished set
for parameter renormalization is given in terms of
$e,M_Z,M_W,M_H,m_f$ with the masses of the corresponding particles.
The finite parts of the counter terms are fixed by the
renormalization conditions that the  propagators have poles
at their physical masses, and $e$ becomes the $ee\gamma$ coupling
constant in the Thomson limit of Compton scattering.
This electroweak on-shell scheme, the 
 extension of the familiar QED renormalization,
has been used in many practical applications
\cite{pav}$^{-}$\cite{LEP}.
 The mass of the Higgs boson,
as long as it is  experimentally unknown,
is treated as a free input parameter.
In practical calculations, the $W$ mass is replaced by $\Gmu$
as an input parameter by using relation (\ref{mw}). 

\smallskip 
Instead of the set  $e,\,M_W,\,M_Z$ as  basic free
parameters other renormalization schemes make use of
$\al$, $\Gmu$,
$M_Z$~\cite{pittau}
 or perform the loop
calculations in the $\overline{MS}$ 
scheme~\cite{pavelt}$^{-}$\cite{msbar2}.
Other schemes applied in the past utilize the parameters
$\al$, $\Gmu$, $\siw$, with the mixing angle deduced
from neutrino-electron scattering~\cite{green}, or the 
concept of effective running couplings~\cite{star,star1}.

\smallskip
Before predictions can be made from the theory,
a set of independent parameters has to be taken from experiment.
For practical calculations the physical input quantities
$ \al, \; \Gmu,\; M_Z,\; m_f,\; M_H, \; \al_s $
are commonly used    
to fix the free parameters of the standard model.
 Differences between various schemes are formally
of higher order than the one under consideration.
 The study of the
scheme dependence of the perturbative results, after improvement by
resummation of the leading terms, allows us to estimate the missing
higher-order contributions (see e.g.\ \cite{yb95} for a comprehensive
study). 

\bigskip
Related to charge and mass renormalization, there occur
two sizeable effects in the electroweak loops 
that deserve a special discussion:

\smallskip \noi
{\it (i) 
Charge renormalization and light fermion contribution:}

\smallskip \noi
Charge renormalization introduces the concept of electric
charge for real photons $(q^2=0)$ to be used for the
calculation of observables at the electroweak scale set
by $M_Z$. Hence the difference 
\beq
 \real \, \Pigr(\mz))  =  \real\, \Pi^{\g}(M_Z^2) - \Pi^{\g}(0)
\eeq
of the photon vacuum polarization is a basic entry in the
predictions for electroweak precision observables.
The purely fermionic contributions correspond to standard QED
and do not depend on the details of the electroweak theory.
They are conveniently split into a leptonic and a hadronic 
contribution
\beq
 \real\,\Pigr (M_Z^2)_{\rm ferm}  =      
                       \real\, \Pigr_{\rm lept}(M_Z^2) \\
                  + \real\, \Pigr_{\rm had}(M_Z^2) \, , 
\eeq
where the top quark is not included in the hadronic
part (5 light flavours); it 
yields a small non-logarithmic contribution
\beq
 \Pigr_{\rm top}(M_Z^2) \simeq \frac{\al}{\pi} \,Q_t^2\,
            \frac{M_Z^2}{5\,m_t^2}  
         \simeq 0.57\cdot 10^{-4}\, .
\eeq 
The quantity
\bea
 \dal & = & \dal_{\rm lept} \, + \,\dal_{\rm had} \nn \\
  & = &   -\,  \real\,\Pigr_{\rm lept}(\mz)
           - \, \real\,\Pigr_{\rm had}(\mz)
\eea
corresponds to a QED-induced shift
in the electromagnetic fine structure constant
\beq
\label{delalpha}
     \al \, \ra \, \al(1+\dal) \, , 
\eeq
which can be resummed according to the renormalization group
accommodating all the leading logarithms of the type
  $ \al^n\log^n(M_Z/m_f)$.
The result can be interpreted as 
an effective fine structure
constant at the $Z$ mass scale:
\beq
\label{alphaeff}
   \al(\mz) \, =\, \frac{\al}{1-\dal} \, .
\eeq
It corresponds to a resummation of the iterated 1-loop vacuum
polarization from the light fermions  to all orders.

$\dal$ is an input of crucial importance because of its universality
and of its remarkable size of $\sim 6 \%$.
The leptonic content can be directly evaluated in terms of the known
lepton masses, yielding at one loop order: 
\beq
 \dal_{\rm lept} = \sum_{\ell=e,\m,\tau}\,
\frac{\al}{3\pi}\left(\log\frac{M_Z^2}{m_{\ell}^2}-\frac{5}{3}
 \right) \, + \, O\left(\frac{m_{\ell}^2}{\mz} \right) .
\eeq
The 2-loop correction has been known already  for a long time
\cite{kallen}, and also the 3-loop contribution is now
available~\cite{steinhauser1}, yielding altogether
\beq
\begin{array}{l}
  \dal_{\rm lept} \; = \; 314.97687\cdot 10^{-4} \; =  \\[0.2cm]
   \left[ 314.19007_{\rm 1-loop} + 0.77617_{\rm 2-loop} 
       + 0.0106_{\rm 3-loop} \right] \cdot 10^{-4} \, .
\end{array}
\eeq

\smallskip \noi
For the light hadronic part, perturbative QCD is not applicable
and quark masses are not available as
reasonable input parameters. Instead,
the 5-flavour contribution to
$\Pigr_{\rm had}$ can be derived from experimental data with the help
of a dispersion relation
\beq
\label{dispersion}
 \dal_{\rm had} = \, -\, \frac{\al}{3\pi} \, M_Z^2 \, \real \,
 \int^{\infty}_{4m_{\pi}^2}  {\rm d}s' \, \frac{R^{\g}(s')}
 {s'(s'-M_Z^2-i\veps)}
\eeq
with
\[
 R^{\g}(s) = \frac{\sigma(\epm \ra\g^* \ra {\rm hadrons})}
 {\sigma(\epm\ra\g^*\ra \m^+\m^-)}
\]
as an experimental input quantity 
in the problematic low energy range.

\begin{figure}[hb]
\centerline{
\epsfig{figure=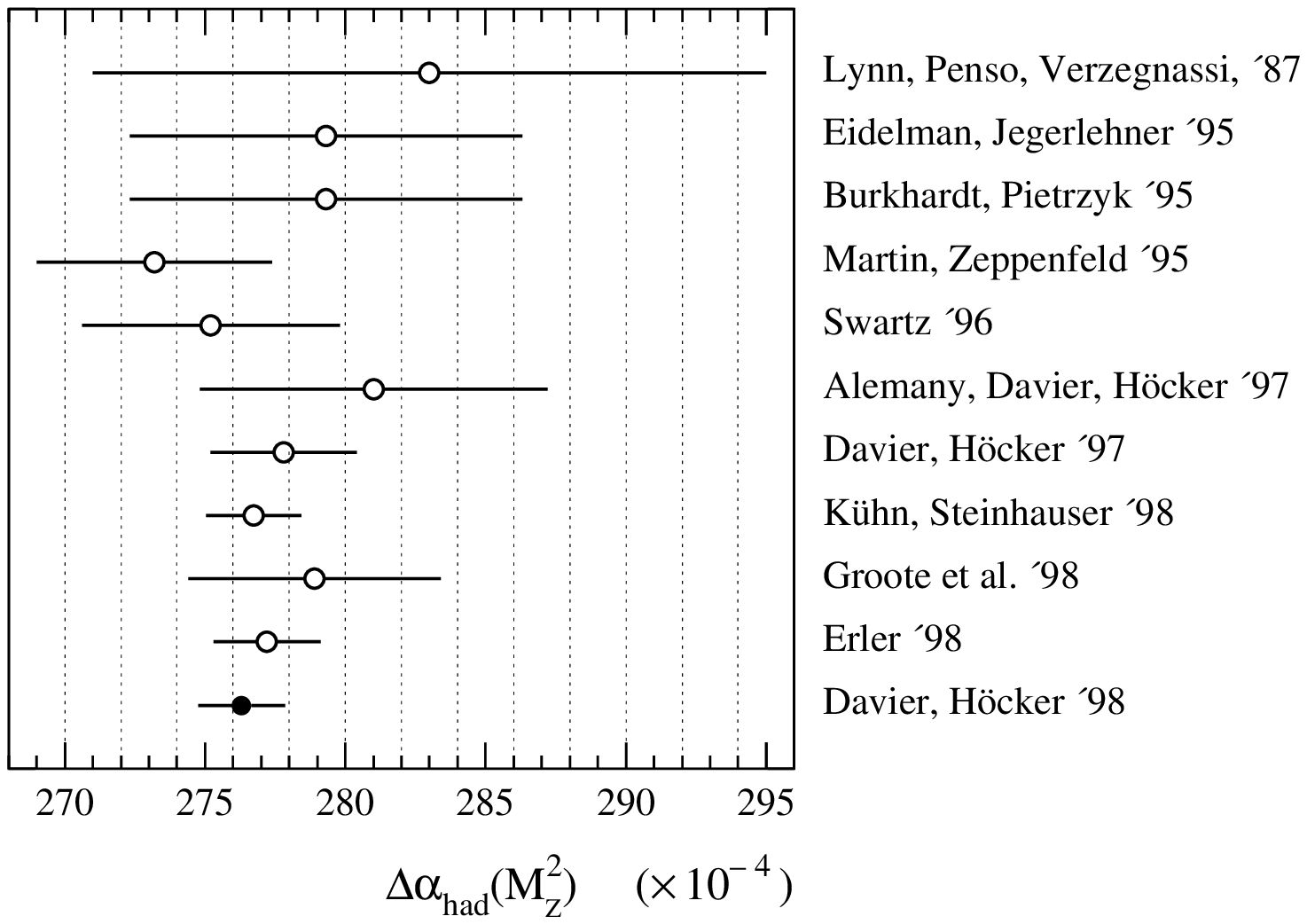,height=6cm}}
\caption{Various determinations of $\dal_{\rm had}$
         (from ref {\protect\cite{hoecker}}). }  
\label{alpha}
\end{figure}

Integrating by means of the trapezoidal rule 
(averaging data in bins) 
over
$\epm$ data for  the energy range below 40 GeV and applying 
perturbative QCD for the high-energy region above, the 
expression (\ref{dispersion}) yields the value
\cite{eidelman,burkhardt}
\beq
\label{fred}
\dal_{\rm had} = - 0.0280\pm 0.0007 \, ,
\eeq
which agrees with another independent analysis
\cite{swartz} with a different error treatment.
Because of the lack of precision in the experimental data 
a large uncertainty 
is associated with the value of $\dal_{\rm had}$,
which propagates into the theoretical error of the predictions
of electroweak precision observables.
Including additional data from $\tau$-decays \cite{alemany}
yields about the same result with a slightly improved
uncertainty.
Recently other attempts have been made to increase  the 
precision of $\dal$ 
\cite{davier,kuehnsteinh}$^{-}$\cite{erler}
by ``theory-driven'' analyses of 
the dispersion integral (\ref{dispersion}).
The common basis is the application of perturbative QCD 
down to the energy scale given by the $\tau$ mass
for the calculation of   
the quantity $R^{\gamma}(s)$ outside the resonances.
Those calculations were made possible  by the recent
availability 
of the quark-mass-dependent $O(\al_s^2)$ QCD corrections
\cite{mqcd} for the cross section down to close to the 
thresholds for $b$ and $c$ production.
[A first step in this direction was done in~\cite{martinzepp}
in the massless approximation.]
In order to pin down the error, two different strategies 
are in use: the application of the method developed in
\cite{mainz} for minimizing  the impact of data from
less reliable regions,
done in~\cite{davier}, and the rescaling of data
in the open charm region of 3.7--5 GeV from PLUTO/DASP/MARKII,
for the purpose of normalization
to agree with perturbative QCD, done in~\cite{kuehnsteinh}.
The results obtained for $\dal_{\rm had}$
are very similar: 
\begin{center}
\btab{l l}
 $ 0.02763\pm 0.0016$ & ref {\protect\cite{davier}} \\
 $ 0.02777\pm 0.0017$ & ref {\protect\cite{kuehnsteinh}} 
\etab
\end{center}
In \cite{erler} the $\ms$ quantity $\hat{\al}(M_Z)$ has been derived
with the help of an unsubtracted dispersion relation in the 
$\ms$-scheme, yielding a comparable error.
The history of the determination of the hadronic vacuum
polarization is visualized in Figure \ref{alpha}.
\vspace{0.5cm}  

The basic assumption in the theory-driven approach, the 
validity of perturbative QCD and quark-hadron duality, 
is supported by the following empirical observations:

-- The strong coupling constant $\al_s(m_{\tau})$ 
determined from hadronic $\tau$ decays shows good agreement
with $\al_s(M_Z)$ determined from $Z$-peak observables
when the renormalization group evolution of $\al_s$
in perturbative QCD is imposed to run $\al_s$ from $m_{\tau}$
to the $Z$-mass scale.

-- Non-per\-tur\-bative con\-tribu\-tions in $R^{\g}(s)$,  
para\-met\-rized in
terms of con\-dens\-ates of quarks, gluons and of vacuum expectation
values of higher-dimensional operators in the operator
product expansion \cite{braaten}
can be probed by comparing spectral moments of $R_{\rm exp}^{\g}(s)$
with the corresponding expressions involving the theoretical
$R^{\g}$. It has been
shown from fitting a set of moments that the non-perturbative
contributions are negligibly small \cite{davier,hoecker}.

--  Recent preliminary measurements of $R^{\g}$ at BES 
at 2.6 and 3.3 GeV show values slightly lower than the
previous data \cite{zhan,karlen}, 
better in alignment with the expectations from
perturbative QCD. 

Although the error in the QCD-based evaluation of
$\dal_{\rm had}$ is considerably reduced, it should be kept in
mind that the conservative estimate in Eq.~(\ref{fred}) is independent
of theoretical assumptions on QCD at lower energies
and thus less sensitive to potential  systematic effects 
not under consideration now \cite{fredjeger}.

\bigskip \noi
{\it (ii) 
Mixing angle renormalization and the $\rho$-parameter:}

\smallskip \noi
The $\rho$-parameter, originally 
defined as the ratio of the neutral to the charged
current strength in neutrino 
scattering$\;$\cite{rho}, 
is unity in the \sm at the
tree level, but gets a deviation $\dro$
from 1 by radiative corrections.
The dominating universal part has its origin in the renormalization
of the relation between the gauge boson masses and the 
electroweak mixing angle. This relation is modified in higher
orders according to 
\bea
  \sin^2\theta_W  = 
   1-\frac{\mw}{\mz} + \frac{\mw}{\mz} \dro\, +  \cdots
\eea
The main contribution to the universal $\rho$-parameter 
\beq
 \rho = \frac{1}{1-\dro}
\eeq
 is from the  $(t,b)$ doublet \cite{rho1},
at the present level calculated as follows:
 \beq
 \label{dro}
 \dro= 3 x_t \cdot [ 1+ x_t \,  \roro+ \drqcd ]
\eeq
with
\beq
\label{xt}
 x_t =
 \frac{\Gmu m_t^2}{8\pi^2\sqrt{2}} \, .
\eeq
 The electroweak 2-loop
 part \cite{bij,barbieri} is described by the
function $\roro(M_H/m_t)$, and
$\drqcd$ is the QCD correction
to the leading $\Gmu m_t^2$ term
 \cite{djouadi,tarasov}
\beq
\label{dro2}
    \delta\rho_{\rm QCD} = -\,
\frac{\al_s(\mu)}{\pi}\, c_1
 +\left(\frac{\al_s(\mu)}{\pi}\right)^2 c_2(\mu)
\eeq
with
$$
 c_1 =  \frac{2}{3} \left(
\frac{\pi^2}{3}+1\right) = 2.8599 
$$
and the  
3-loop coefficient$\;$\cite{tarasov} 
$c_2(\mu)$, which amounts to
$$
 c_2=-14.59 \; 
 \mbox{  for } \mu =m_t \mbox{ and 6 flavours} 
$$
with the on-shell top mass $m_t$.
This reduces the scale
dependence of $\rho$ significantly and hence is an important
entry to decrease the theoretical uncertainty of the standard model
predictions for precision observables.

\smallskip 
There is also a Higgs contribution to $\dro$,  which, however, is not
UV-finite by itself when derived from only the diagrams involving the
physical Higgs boson.
The $M_H$-dependence for large
Higgs masses $M_H$ is
 only logarithmic in 1-loop order \cite{screening};
the 2-loop contribution \cite{bij1} shows a dependence
$\sim M_H^2$ for large values of the Higgs mass. 
In the limit
    $\sin^2\theta_W \ra 0, M_Z\ra M_W$, where the $U(1)_Y$ is switched
off, one finds $\dro_{\rm H} = 0$. This is the consequence
of the global $SU(2)_R$ symmetry of the Higgs Lagrangian
(`custodial symmetry'), which is broken by the $U(1)_Y$ group.
Thus, $\dro_{\rm H}$ is a measure of the $SU(2)_R$ breaking by the
weak hypercharge.


%
\subsection{Muon decay and the vector boson masses}
The interdependence between the gauge boson masses is established
through the accurately measured muon lifetime or, 
equivalently, the Fermi coupling constant $\Gmu$.
Originally, the $\m$-lifetime $\tau_{\m}$ has been calculated
within the framework of the effective 4-point Fermi interaction.
Beyond the well-known 1-loop QED corrections \cite{muon}, 
the 2-loop QED corrections in the Fermi model 
have been calculated quite recently \cite{stuartvanritbergen},
yielding the expression
(the error in the 2-loop term is from the hadronic uncertainty)
\bea
\label{taumu}
\frac{1}{\tau_{\m}} & = & \frac{\Gmu^2m_{\m}^5}{192\pi^3}
\left( 1-\frac{8m_e^2}{m_{\m}^2} \right) \cdot  \\[0.1cm]
 & \cdot &  \left[ 1+1.810\, \alpi+(6.701\,\pm 0.002)
       \left(\alpi\right)^2 \right] \, . \nn
\eea
This formula is the defining equation for $\Gmu$ in terms
of the experimental $\m$-lifetime. 
Owing to the presence of order-dependent QED corrections, the
numerical value of the Fermi constant changes after
the second-order term is included. 
Compared with the value given in the 1998
report of the Particle Data Group \cite{PDG98}, the
latest value is now smaller by $2\cdot 10^{-10}$ $\gv^{-2}$, 
namely \cite{stuartvanritbergen}
\beq
   \Gmu = (1.16637\pm 0.00001) \cdot 10^{-5} \, \gv^{-2}\, , 
\eeq
where also the error has been reduced by a factor 
of about 1/2.

In the standard model, $\Gmu$ can be calculated including quantum
corrections in terms of the basic \sm parameters, thereby
separating off all diagrams that correspond to the QED
corrections in the Fermi model. This yields 
the correlation between
the masses $M_W,M_Z$ of the vector bosons, expressed in terms
of $\al$ and $\Gmu$;  in 1-loop order it is given by
 \cite{sirmar}:
\beq
\label{mw}
\frac{\Gmu}{\sqrt{2}}   =
            \frac{\pi\al}{2s_W^2 M_W^2} [
        1+ \Dr(\al,M_W,M_Z,M_H,m_t) ] \, .
\eeq
with 
$ \quad s_W^2 = 1- \mw/\mz $ .

\smallskip \noi
The decomposition
\beq
 \Dr = \Delta\al -\frac{c_W^2}{s_W^2}\,\dro^{(1)}
         + (\Dr)_{\rm rem} 
\eeq
separates the
leading fermionic contributions $\dal$ and $\dro$(1-loop).
All other terms are collected in the remainder part 
$(\Dr)_{\rm rem}$,
the typical size of which is of order $\sim 0.01$.
 
\bigskip
The presence of large terms in $\Dr$ requires the consideration
of effects higher than 1-loop (see also the contribution by K\"uhn
\cite{kuehntalk} to these proceedings).
The modification of Eq.\ (\ref{mw}) according to
\bea
\label{resum}
         1+\Dr & \, \ra\,& \frac{1}{(1-\Delta\al)\cdot
(1+\frac{c_W^2}{s_W^2}\dro) \, -\,(\Dr)_{\rm rem}}
 \nn \\
 & \equiv & \frac{1}{1-\Dr}
\eea
accommodates the following higher-order terms
($\Dr$ in the denominator is an effective correction including
higher orders):

\smallskip 
(i) \hspace{0.1cm} 
the leading log resummation \cite{marciano} of $\dal$:
$  1+\dal\, \ra \, (1-\dal)^{-1}$ \, ;

\smallskip
(ii) \hspace{0.2cm}
the resummation of the leading $m_t^2$ contribution \cite{chj}
in terms of $\dro$ in Eq.\ (\ref{dro}).
Beyond the QCD higher-order contributions through the $\rho$-parameter,
the complete
 $O(\al\al_s)$ corrections to the self energies
 are available 
\cite{qcd,dispersion1}.
All these higher-order terms contribute with the same positive sign
to $\Dr$.
Non-leading QCD corrections  to $\Delta r$ of
the type
$$ \Dr_{(bt)} = 3x_t \left(\frac{\al_s}{\pi}\right)^2 \left(
 a_1 \frac{\mz}{m_t^2} + a_2 \frac{M_Z^4}{m_t^4} \right)
$$
are also available \cite{steinhauser}.

\smallskip
(iii) \hspace{0.1cm}
With the quantity $(\Dr)_{\rm rem}$ in the denominator,
non-leading higher-order terms
containing mass singularities of the type $\al^2\log(M_Z/m_f)$
from light fermions
are incorporated \cite{nonleading}.

\smallskip
(iv) \hspace{0.2cm}
The subleading $\Gmu^2m_t^2 M_Z^2$ contribution 
of the electroweak 2-loop order \cite{padova}
in an expansion in terms of the top mass.
This subleading term turned out to be sizeable, about as large
as the formally leading term of $O(m_t^4)$ via the $\rho$-parameter.
In view of the present and future experimental accuracy it  
constitutes a non-negligible shift in the $W$ mass.

\smallskip 
Meanwhile
exact results have been derived for the 
Higgs-dependence of the fermionic 2-loop corrections in $\Dr$
\cite{bauberger}, and comparisons were performed with
those obtained via the top mass expansion \cite{weiglein}.
Differences in the values of $M_W$ 
of several MeV (up to 8 MeV) are observed when $M_H$ is varied
over the range from 65 GeV to 1 TeV.

\begin{figure}[htb]
\centerline{
\epsfig{figure=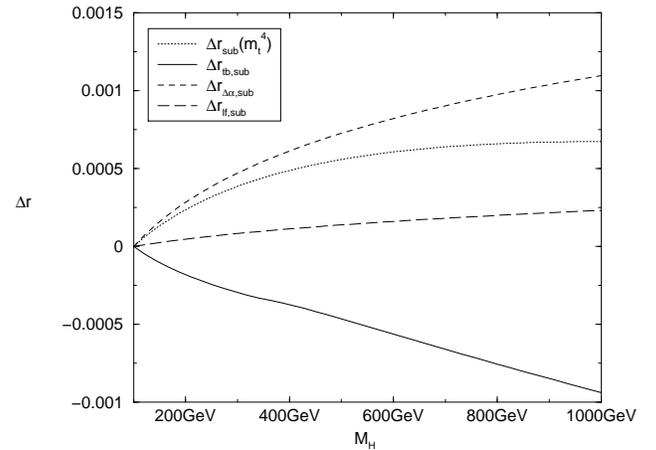,height=6cm}}
\caption{Higgs mass dependence of fermionic contributions 
         to $\Dr$ at the two-loop level 
         (from {\protect\cite{weiglein}}). 
         The different curves
         show the various contributions:
         light fermions via $\dal$ ($\Dr_{\dal}$),
         residual light-fermion contribution not in $\dal$
         ($\Dr_{\rm lf}$), the 
         contribution from the $(t b$) doublet ($\Dr_{\rm tb}$),
         and the approximation of the $(tb)$ two-loop 
         contribution by the term proportional to $m_t^4$.
         Displayed
         in each case is the difference
         $\Dr(M_H) - \Dr(100$ GeV). }
\label{achim_deltar}
\end{figure}

Figure \ref{achim_deltar} shows the Higgs-mass dependence
of the two-loop corrections to $\Dr$ associated with the
$t/b$ doublet, with $\dal$, and with the light fermion terms
not in $\dal$, together with the leading $m_t^4$-term, which
constitutes a very poor approximation. 

Pure fermion-loop contributions ($n$ fermion loops at
$n$-loop order) have also been investigated 
\cite{weiglein,achim1}. In the on-shell scheme,
explicit results have been worked out up to 4-loop
order, which allows an investigation of the validity of the
resummation (\ref{resum})
for the non-leading 2-loop and higher-order
terms. It was found that numerically the resummation 
(\ref{resum}) works remarkably well, within 2 MeV in $M_W$.

\subsection{$Z$ boson observables}
Measurements
of the $Z$ line shape in $\epmf$
yield the parameters
$M_Z,\, \G_Z$,   and the partial widths $\G_f$ or the peak
cross section
\beq
\sigma_0^f = \frac{12\pi}{\mz}\cdot\frac{\G_e\G_f}{\G_Z^2} \, .
\eeq
Angular distributions and polarization measurements of the final
fermions yield forward--backward and polarization asymmetries.
Whereas $M_Z$ is used as a precise input parameter, together
with $\al$ and $\Gmu$, the width, partial widths
and asymmetries allow
comparisons with the predictions of the standard model.
The predictions for the partial widths
as well as for the asymmetries
can conveniently be calculated in terms of effective neutral
current coupling constants for the various fermions.
 
\paragraph{\it Effective $Z$ boson couplings:}
 
The effective couplings follow
from the set of 1-loop diagrams
without virtual photons,
the non-QED  or weak  corrections.
These weak corrections
can conveniently be written
in terms of fermion-dependent overall normalizations
$\rho_f$ and effective mixing angles $s_f^2$
in the NC vertices (see e.g.~\cite{ewgr}):
\bea
\label{nccoup}
 & &
 J_{\nu}^{\rm NC}  =   \left( \sqrt{2}\Gmu\mz \right)^{1/2} \,
  (g_V^f \,\ganu -  g_A^f \,\ganu\gafi)  \\
 & &
   =  \left( \sqrt{2}\Gmu\mz \rho_f \right)^{1/2}
\left( (I_3^f-2Q_fs_f^2)\ganu-I_3^f\ganu\gafi \right)  . \nn
\eea
$\rho_f$ and $s_f^2$ contain  universal
parts, e.g.\ from the $\rho$-parameter via 
\beq
\rho_f  =  \frac{1}{1-\dro} + \cdots , \;\;\;
 s_f^2  = s_W^2 + c_W^2\,\dro + \cdots
\eeq
with $\dro$ from Eq.\ (\ref{dro}) and
non-universal parts that explicitly depend on the type of the
external fermions.

The
subleading 2-loop corrections
$\sim \Gmu^2m_t^2 M_Z^2$  
for the leptonic mixing angle \cite{padova}
$s_{\ell}^2$  
have also been obtained in the meantime, as well as 
for $\rho_{\ell}$ \cite{padova1}. 

Meanwhile
exact results have been derived for the 
Higgs-dependence of the fermionic 2-loop corrections in $\sell$
\cite{weiglein,achim1}, and comparisons were performed with
those obtained via the top mass expansion \cite{weiglein}.
Differences in the values of $\sell$ 
can amount to $0.8\cdot 10^{-4}$ when $M_H$ is varied
over the range from 100 GeV to 1 TeV.

\begin{figure}[htb]
\centerline{
\epsfig{figure=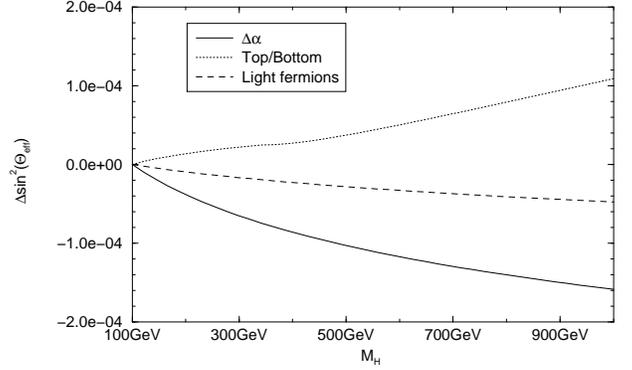,height=5cm}}
\caption{Higgs mass dependence of the
         various fermionic contributions at the two-loop  
         level to the effective
         leptonic mixing angle $\sell$ at the $Z$ peak: 
         light fermion contribution via $\dal$,
         light fermion contribution  not in $\dal$,
         and the contribution from the $(t b)$ doublet.
         Shown in each case is the difference
         $\sell(M_H)-\sell(100\,$GeV). }
\label{achim_sin}
\end{figure}

Figure \ref{achim_sin} shows the Higgs-mass dependence
of the 2-loop corrections to $\sell$ associated with the
$t/b$ doublet, with $\dal$, and with the light fermion terms
not in $\dal$.
As can be seen, the $M_H$-dependence of the light fermions 
yields contributions to $\sell$ up to $2\cdot 10^{-5}$.
For $\rho_{\ell}$ or equivalently the leptonic $Z$ widths, 
the subleading 2-loop effects are small, and differences with the
results in the top mass dependence are irrelevant.
 
\smallskip
For the $b$ quark coupling to the $Z$ boson, not only the
universal contribution through the $\rho$-parameter but 
also the non-universal parts have a strong
dependence on $m_t$, resulting from virtual top quarks in the
vertex corrections. The difference between the $d$ and $b$
couplings can be parametrized in the following way
\beq
  \rho_b = \rho_d (1+\tau)^2, \;\;\;\;
  s^2_b = s^2_d (1+\tau)^{-1} \, ,
\eeq
with the quantity
$$
 \tau = \Delta\tau^{(1)}
      + \Delta\tau^{(2)}
      + \Delta\tau^{(\al_s)}
$$
calculated perturbatively, including
the complete 1-loop order term \cite{vertex} with $x_t$
from Eq.\ (\ref{xt}): 
\beq
\Delta\tau^{(1)} = -2 x_t - \frac{\Gmu\mz}{6\pi^2\sqrt{2}}
 (c_W^2+1)\log\frac{m_t}{M_W} + \cdots ,
\eeq
and  the leading
electroweak 2-loop contribution of $O(\Gmu^2 m_t^4)$
\cite{barbieri,dhl}
\beq
\Delta\tau^{(2)} = -2\, x_t^2 \, \tau^{(2)} \, ,
\eeq
where
 $\tau^{(2)}$ is a function of $M_H/m_t$
with
 $\tau^{(2)} = 9-\pi^2/3$ for small $M_H$.
 
\smallskip
\paragraph{\it Asymmetries and mixing angles:}
 
The effective mixing angles are of particular interest, since
they determine the on-resonance asymmetries via the combinations
   \beq
    A_f = \frac{2g_V^f g_A^f}{(g_V^f)^2+(g_A^f)^2}  \, ,
\eeq
namely
\beq
 \afb =\frac{3}{4}\, A_e A_f, \quad A_{\tau}^{\rm pol} = A_{\tau},
 \quad \alr = A_e \, .
\eeq
Measurements of the \ass hence are measurements of
the ratios
\beq
  g_V^f/g_A^f = 1 - 2 Q_f s_f^2
\eeq
or the effective mixing angles, respectively.

\smallskip
\paragraph{\it $Z$ width and partial widths:}
 
The total
$Z$ width $\Gamma_Z$ can be calculated
essentially as the sum over the fermionic partial decay widths.
Expressed in terms of the effective coupling constants,
they read up to second order in the fermion masses:
\bea
\Gamma_f
  & = & \G_0
 \left[ 
     (g_V^f)^2  +
     (g_A^f)^2 \left(1-\frac{6m_f^2}{\mz} \right) \right]
 \cdot   \left( 1+ Q_f^2\, \frac{3\al}{4\pi} \right) \nn \\
     & +& \Delta\G^f_{\rm QCD} \nn
\eea
with
$$
\G_0 \, =\,
  N_C^f\,\frac{\sqrt{2}\Gmu M_Z^3}{12\pi},
 \;\;\;\; N_C^f = 1
 \mbox{ (leptons)}, \;\; = 3 \mbox{ (quarks)}.
$$
The QCD correction for the light quarks
with $m_q\simeq 0$ is given by
\beq
\label{QCD}
 \Delta\G^f_{\rm QCD}\, =\, \G_0
  \left( (g_V^f) ^2+ (g_A^f)^2 \right)
 \cdot K_{\rm QCD}
\eeq
with \cite{qcdq}
$$
K_{\rm QCD}  =   \frac{\al_s}{\pi} +1.41 \left(
  \frac{\al_s}{\pi}\right)^2 -12.8 \left(
  \frac{\al_s}{\pi}\right)^3
  -\frac{Q_f^2}{4}\frac{\al\al_s}{\pi^2} \,  .  
$$
For $b$ quarks
the QCD corrections are different, because of finite $b$ mass terms
and to top-quark-dependent 2-loop diagrams
 for the axial part:
\beq
 \Delta\G_{\rm QCD}^b  =
 \Delta\G_{\rm QCD}^d \, +\, \G_0 \left[
           (g_V^b)^2  \, R_V \,+\,
           (g_A^b)^2 \, R_A  \right]  \, . \nn
\eeq
The coefficients in the perturbative expansions
\bea
 R_V &=& c_1^V \aspi + c_2^V \left(\aspi\right)^2 
          + c_3^V \left(\aspi\right)^3 + \cdots,
 \nn \\
 R_A &=&  c_1^A \aspi + c_2^A \left(\aspi\right)^2 + \cdots \,
 \nn
\eea
depending on $m_b$ and $m_t$,
are calculated up to third order in $\al_s$, except for the 
$m_b$-dependent singlet terms,
which are known to $O(\al_s^2)$ \cite{qcdb,qcdb1}. 
For a review of the QCD
corrections to the $Z$ width, see \cite{qcdq1}.

The partial decay rate into $b$-quarks, in particular the
ratio $R_b = \G_b/\G_{\rm had}$, is an observable of
special sensitivity to the top quark mass. Therefore, 
beyond the pure QCD corrections, also the 2-loop contributions
of the mixed QCD--electroweak type, are important.
The QCD corrections were first derived for 
the leading term of $O(\al_s\Gmu m_t^2)$
\cite{jeg}
and were subsequently completed by the 
 $O(\al_s)$ correction to the $\log m_t/M_W$ term 
\cite{log} and the residual terms of $O(\al\al_s)$ \cite{harlander}.

In the same spirit, also the complete 2-loop 
$O(\al\al_s)$ to the partial widths into the light quarks 
have been obtained, beyond those that are already contained in
the factorized expression (\ref{QCD}) with the electroweak
1-loop couplings \cite{czarnecki}. These 
``non-factorizable'' corrections  
yield an extra negative contribution of 
$-0.55(3)$ MeV to the total hadronic $Z$ width
(converted into a shift of the strong coupling constant, they
correspond to $\delta \al_s = 0.001$).
In summary, the 2-loop corrections of $O(\al\al_s)$ 
to the electroweak precision observables are
by now completely under control.  
More details can be found in \cite{kuehntalk}.

Radiation of secondary fermions through photons
from the primary final state fermions
can yield another sizeable contribution to the partial $Z$ widths;
however, this is compensated by the corresponding virtual
contribution through the dressed photon propagator in the final-state
vertex correction for sufficiently inclusive final states,
i.e.\ for loose cuts to the invariant mass
of the secondary fermions \cite{hoang}.

\paragraph{\it QED corrections:}
 The observed cross section is the result of convoluting
the cross section for $\epmf$
calculated on the basis of the
effective couplings  
with the initial-state QED corrections consisting
of virtual photon and real photon bremsstrahlung
contributions:
\beq
\label{convolution}
\sigma_{\rm obs}(s)  =
 \int_0^{k_{\rm max}}\, dk \, H(k)\, \sigma(s(1-k)) \, ;
\eeq
$k_{\rm max}$ denotes a cut to the radiated energy. Kinematically
it is limited by $1-4m_f^2/s$ or
$1-4m_{\pi}^2/s$ for hadrons, respectively.
For the required accuracy, multiphoton radiation has to be
included. The radiator function $H(k)$ with soft-photon
resummation and the exact $O(\al^2)$ result
 for initial-state
QED corrections is given in ref~\cite{berends}.
It has been improved recently  
by the $O(\al^3)$ term  
 \cite{montagna}.

Bhabha scattering in the forward direction is the crucial
theoretical tool for the determination of the luminosity 
and therefore requires a careful treatment, including 
higher-order QED corrections \cite{bhabha}.
Improvements in the calculation of the $O(\al^2)$
next-to-leading logarithmic contributions in the Monte Carlo 
generator BHLUMI are an important
step in pinning down the theoretical error from 0.11\%
to 0.06\% \cite{bennie}.

\subsection{Accuracy of the standard model predictions}
 For a discussion of the theoretical reliability
of the \sm predictions, one has to consider the various sources
contributing to their
uncertainties:

{\it Parametric uncertainties} result from the limited 
precision in the experimental values of the input
parameters, essentially $\al_s = 0.119\pm 0.002$ \cite{PDG98},
$m_t=173.8\pm 5.0$ GeV \cite{partridge},
$m_b=4.7\pm 0.2$ GeV,
 and the
hadronic vacuum polarization as discussed in section 2.1.
The conservative estimate of the
error in Eq.~(\ref{fred})
leads to
$\delta M_W = 13$ MeV in the $W$-mass prediction, and
$\delta\sin^2\theta = 0.00023$ common to all of the mixing
angles.

The uncertainties from the QCD contributions
can essentially be traced back to
those in the top quark loops in the vector boson self-energies.
The knowledge of the $O(\al_s^2)$ corrections to the
$\rho$-parameter and $\Dr$ yields a significant reduction;
they are small, although not negligible (e.g. 
$\sim 3\cdot 10^{-5}$ in $\sell$).

The size of unknown higher-order contributions can be estimated
by different treatments of non-leading terms
of higher order in the implementation of radiative corrections in
electroweak observables (`options')
and by investigations of the scheme dependence.
Explicit comparisons between the results of 5 different computer codes  
based on  on-shell and $\ms$ calculations
for the $Z$-resonance observables are documented in the ``Electroweak
Working Group Report'' \cite{ewgr} in ref \cite{yb95}.
The inclusion of the 
non-leading 2-loop corrections
$\sim \Gmu^2m_t^2 M_Z^2$ 
reduce the uncertainty in $M_W$ below 10 MeV  
and in $s^2_{\ell}$ below $10^{-4}$, typically
to $\pm 4\cdot 10^{-5}$.

\section{Standard model and precision data}
We now confront the standard model predictions for
the discussed set of precision observables 
with the most 
recent sample of experimental data \cite{lep,karlen}.
In table \ref{zobs}
the \sm predictions for $Z$-pole observables and the $W$ mass  are
put together for the best fit input data set, given in
(\ref{bestfit}). 
The experimental results on the $Z$ observables are from 
LEP and the SLC, 
the $W$ mass is from combined LEP and $p\bar{p}$ data.
The leptonic mixing angle determined via $\alr$ by the SLD
experiment~\cite{sld} and the 
$s^2_{\ell}$ average 
from LEP:
\bea
   s^2_e(\alr) & = & 0.23109 \pm 0.00029  \nn \\
   s^2_{\ell} ({\rm LEP}) & = & 0.23189 \pm 0.00024  \nn
\eea
have come closer to each other in their central value; owing to their
smaller errors, however, they still differ by 2.8 standard deviations.

\begin{table}[t]
\caption[]   {\label{zobs}Precision observables: 
              experimental results from combined
              LEP and SLD data for $Z$ observables and 
              combined $p\bar{p}$
              and LEP data for $M_W$, 
             together with the
             standard model         
             predictions for the best fit, i.e.
             for the parameter values given in Eq.~(\ref{bestfit}). 
             $\rho_{\ell}$ and $s^2_{\ell}$ are derived from the
             experimental values of
             $g_{V,A}^{\ell}$ according to Eq.~(\ref{nccoup}),
             averaged under the assumption of lepton
             universality. }
\vspace{0.5cm}
\begin{center}
 \btab{@{}| l  l  r | }
\hline 
Observable & Exp.  & SM  best fit \\
\hline 
  & &  \\[-0.3cm]
$M_Z$ (GeV) & $91.1867\pm0.0019$ & 91.1865    \\
$\Gamma_Z$ (GeV) & $2.4939\pm 0.0024$ & 2.4956 \\
$\sigma_0^{had}$ (nb) & $41.491\pm 0.058$ & 41.476 \\
 $ R_{had}$ & $20.765\pm 0.026 $ & 20.745 \\
$R_b$  & $0.21656\pm 0.00074$ & 0.2159 \\
$ R_c$  & $0.1732\pm0.0048$ & 0.1722 \\
$\afb^{\ell}$ & $0.01683 \pm 0.00096$ & 0.0162 \\
$\afb^b$ & $0.0990 \pm 0.0021$ &  0.1029 \\
$\afb^c$ & $0.0709 \pm 0.0044$ &  0.0735 \\
$A_b$            & $0.867\pm 0.035$  & 0.9347 \\
$A_c$            & $0.647\pm 0.040$  & 0.6678 \\
$\rho_{\ell}$ & $1.0041\pm 0.0012$ &  1.0051 \\
$s^2_{\ell}$  & $0.23157\pm 0.00018$ & 0.23155\\
$M_W$ (GeV) & $80.39 \pm 0.06$ & 80.372 \\[0.1cm]
\hline
\etab
 \vspace{0.5cm}
\end{center}
\clearpage
\end{table}

Table \ref{zobs} contains the combined LEP/SLD value.
 $\rho_{\ell}$ and $s^2_{\ell}$ are the leptonic
neutral current couplings in Eq.~(\ref{nccoup}), 
derived from partial widths and
asymmetries  under the assumption of lepton universality.

Note that
the experimental value for $\rho_{\ell}$ points at the presence of
genuine electroweak corrections by 3.5 standard deviations. 
In $\sell$  the presence of purely bosonic radiative corrections 
is clearly established when the experimental result is compared
with a theoretical value containing only the fermion loop corrections,
an observation that has been persisting already for several 
years~\cite{gambinosirlin}.
The 
deviation from the \sm prediction in the quantity $R_b$ has been reduced
below   one standard deviation by now.
Other small deviations 
 are observed in the asymmetries: the purely leptonic $\afb$ is slightly
higher than the standard model predictions, and $\afb$ for $b$ quarks is 
lower.
Whereas the leptonic $\afb$ favours a very light Higgs boson, 
the $b$ quark asymmetry needs a heavy Higgs.

\begin{figure}[htb]
\centerline{
\epsfig{figure=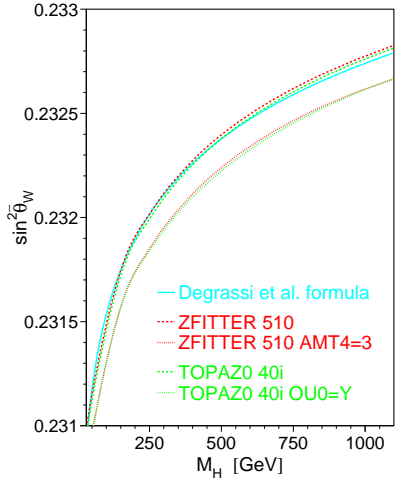,height=8cm}}
\vspace{-0.5cm}
\caption{Higgs mass dependence of $\sell$ with and without
         the electroweak 2-loop term $\sim m_t^2$, 
         comparison of ZFITTER and TOPAZ0 codes. The lower
         sample of curves is without, the upper sample with
         the 2-loop term
         (figure prepared by C. Pauss).}
\label{bardin_sin}
\end{figure}

The effective mixing angle is an observable 
most sensitive to the mass $M_H$ of the Higgs boson.
Since a light Higgs boson  corresponds to
a low value of $s^2_{\ell}$,
the strongest upper bound on $M_H$ is from $\alr$ at the SLC \cite{sld}.
The inclusion of the two-loop electroweak corrections
$\sim m_t^2$ from \cite{padova} yields a sizeable positive
contribution to $\sell$, see Figure \ref{bardin_sin}.
The inclusion of this term hence strengthens the upper bound on $M_H$.

%

\smallskip 
The $W$ mass prediction in table \ref{zobs}
is obtained  from Eq.~(\ref{mw}) 
(including the higher-order terms)
from
 $M_Z,\Gmu,\al$ and  $M_H,m_t$.
The present experimental value for the $W$ mass
from the combined LEP 2, UA2,  CDF and D0 results
is in best agreement with the \sm prediction. 

\smallskip
The quantity $s_W^2$ resp.\  the ratio $M_W/M_Z$
can indirectly be measured in deep-inelastic
neutrino--nucleon scattering.
The average from the experiments
CCFR, CDHS and CHARM \cite{neutrino} 
with the recent NUTEV result \cite{nutev}
\beq
\label{sw}
s_W^2 = 1 - M_W^2/M_Z^2 = 0.2255 \pm 0.0021   
\eeq
for $m_t=175$ GeV and $M_H=150$ GeV
corresponds to $M_W = 80.25 \pm 0.11$ GeV and is hence 
fully consistent with the direct vector boson mass measurements
and with the standard theory.

\begin{figure}[htb]
\centerline{
\epsfig{figure=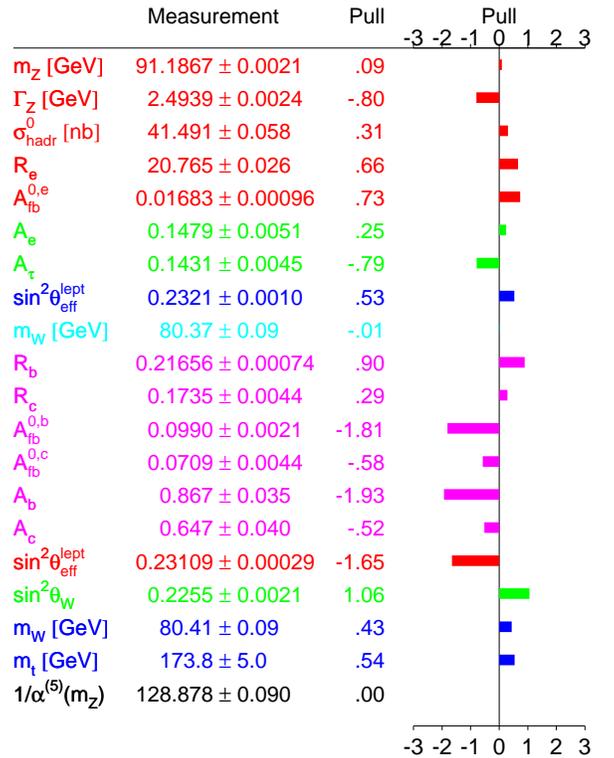,height=12cm}}
\vspace{-1.0cm}
\caption{Experimental results and pulls from a standard model fit
         (from ref {\protect\cite{lep,karlen}}).
         pull = obs(exp)-obs(SM)/(exp.error).} 
\label{pull}
\end{figure}

\paragraph{\it Standard model global fits:}

\begin{figure}[htb]
\centerline{
\epsfig{figure=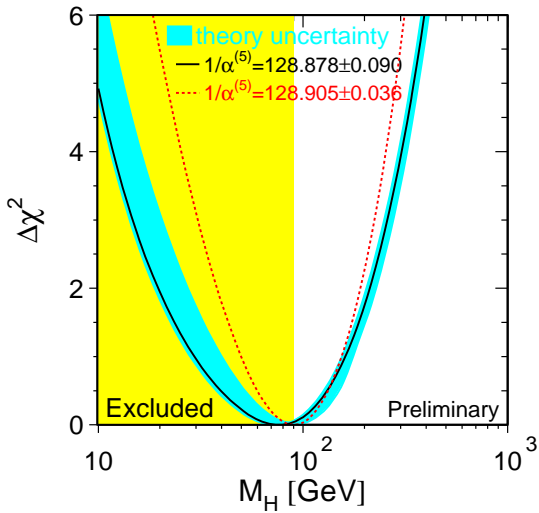,height=8cm}}
\vspace{-0.5cm}
\caption{Higgs mass dependence of $\chi^2$ in the global fit
         to precision data 
         (from ref {\protect\cite{lep,karlen}}).
         The shaded band displays the error from the theoretical
         uncertainties obtained from various options in the
         codes ZFITTER and TOPAZ0. } 
\label{blueband}
\end{figure}

The FORTRAN codes ZFITTER \cite{zfitter} and TOPAZ0 
\cite{topaz0} 
have been updated by incorporating all the recent 
precision calculation results that were 
discussed in the previous section.
Comparisons have shown good agreement between the
predictions from the two independent programs \cite{bardinpassarino}.
Global fits of the \sm parameters to the electroweak precision
data done by the Electroweak Working Group \cite{lep} are based
on these recent versions.
Including $m_t$ and $M_W$ from the direct measurements in the
experimental data set, together with $s_W^2$ from neutrino scattering, 
the \sm parameters for the best fit result are:
\bea
\label{bestfit}
 m_t & =& 171.1 \pm 4.9 \, \gv \nn \\
 M_H & = & 76^{+85}_{-47} \, \gv \nn \\
 \al_s & = & 0.119 \pm 0.003 \, . 
\eea
The upper limit to the Higgs mass at the 95\% C.L. is 
$M_H < 262$ GeV, where the theoretical uncertainty is included.
 Thereby the hadronic vacuum polarization 
in Eq.~(\ref{fred}) has been used (solid line in Figure
\ref{blueband}). 
With the theory-driven
result on $\dal_{\rm had}$ of ref \cite{davier} 
one obtains~\cite{lep} $ M_H=92^{+64}_{-41}$ (
dashed line).
The $1\sigma$ upper bound on $M_H$ is influenced only 
marginally. The reason is that simultaneously with the
error reduction the central value of $M_H$ is shifted 
upwards (see Figure \ref{blueband}). 
Another recent analysis \cite{langacker} 
(for earlier studies see \cite{higgsfits,deboer})
based on the data set of summer 1998
 yields a Higgs mass
$M_H = 107^{+67}_{-45}$ GeV. About one half of the difference
with (\ref{bestfit}) can be ascribed to the use of 
$\al(M_Z)$ of ref \cite{erler}, which is very close to the value
in ref \cite{davier,kuehnsteinh}; the residual shift
might be interpreted as due to different renormalization 
schemes and different treatments of $\al_s$.

With an overall $\chi^2/{\rm d.o.f.} = 15/15$ the quality of
the fit is remarkably high.
As can be seen from Figure \ref{pull}, the deviation of the 
individual quantities from the \sm best-fit values 
are below 2 standard deviations.

\begin{figure}[htb]
\vspace{-1.5cm}
\centerline{
\epsfig{figure=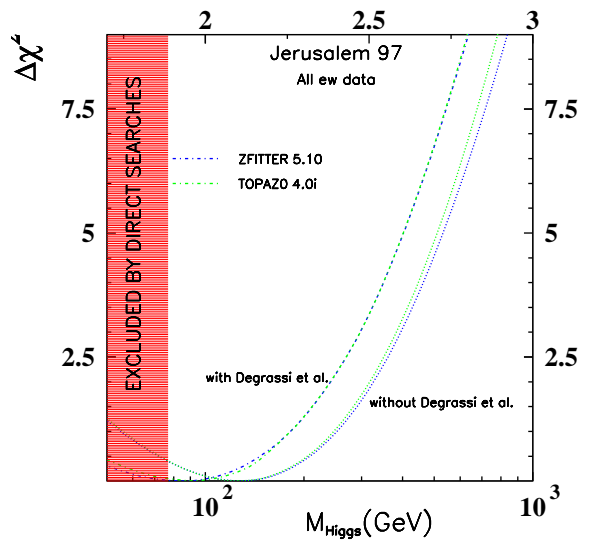,height=12cm}}
\vspace{-2.0cm}
\caption{Higgs mass dependence of $\chi^2$ in the global fit
         to precision data with and without the 2-loop term
         (figure prepared by  G.~Quast). }
\label{bardin_chi2}
\end{figure}

Compared with the results from 1997, the central value for the Higgs
mass has moved to lower values and the error has been decreased.
The Higgs mass bounds follow from the $\chi^2$ distribution
shown in Figure \ref{blueband}. The shift in the central value
 can be understood from Figure \ref{bardin_chi2}, which illustrates
the effect of the inclusion of the electroweak two-loop contribution
by Degrassi et al.\ \cite{padova}, which was not implemented
in the codes for the analysis in 1997. Since it increases the
prediction for $\sell$ (Figure \ref{bardin_sin}) for a given
Higgs mass, the allowed values of $M_H$ are shifted 
accordingly downwards.

The second observation is the decrease of the error, which
besides the experimental improvements results from the 
reduction of the theoretical uncertainties of pure electroweak
origin. The shaded band around the solid line in
Figure \ref{blueband} is the influence of the various `options'
(see section 2.4)
in the codes ZFITTER and TOPAZ0
after the implementation of the 2-loop electroweak terms
$\sim m_t^2$. It is thus the direct continuation of the
error estimate done in the previous study  \cite{ewgr}.
Compared with the width of the uncertainty band in 1997 \cite{lep}
the shrinking is evident.

On the other hand, the remaining theoretical uncertainty 
associated with the Higgs mass
bounds should be taken very seriously. The effect of the 
inclusion of the next-to-leading  
term in the $m_t$-expansion of the electroweak 2-loop
corrections in the precision observables
has shown to be sizeable, 
at the upper margin of the
estimate given in \cite{ewgr}. 
It is thus not guaranteed that the subsequent subleading terms
in the $m_t$-expansion are indeed smaller in size. K\"uhn 
\cite{kuehntalk} has given an example for an explicit 
calculation where the subleading terms  of the $m_t$-expansion
are of comparable size and tend to
cancel each other. Also the variation of the $M_H$-dependence
at different stages of the calculation, as
discussed in sections 2.2 and 2.3, indicate the necessity of
more complete results at two-loop order. 
Having in mind also the
variation of the Higgs mass bounds under the 
fluctuations of the experimental data \cite{karlen},
the limits for $M_H$ derived from the analysis of 
electroweak data  in the frame of the \sm 
still carry a noticeable uncertainty. Nevertheless, 
as a central message, it can be concluded
that the indirect determination of the Higgs mass range 
has shown that the Higgs is light, with its mass well
below the non-perturbative regime.

\section{Production and decay of $W$ bosons}
The success of the \sm in the correct description of the
electroweak precision observables is simultaneously an indirect
confirmation of the Yang--Mills structure of the gauge boson
self-interaction. For conclusive confirmations the direct 
experimental investigation is required. 
At LEP 2 (and higher energies), pair production of on-shell
$W$ bosons can be studied experimentally, allowing tests
of the trilinear vector boson self-couplings and precise
$M_W$ measurements. 
For LEP 2, an error of about 40 MeV in $M_W$
can be reached \cite{kunszt}.
For this purpose  
standard model calculations for the process
$\epm \ra W^+W^- \ra 4 f$ and the corresponding 4-fermion background
processes  are mandatory
at the accuracy level of at least 1\%.
This requires the understanding of the radiative corrections to
the $W$ boson production and decay processes, as well as a careful
treatment of the finite-widths effects.  

For practical purposes, improved Born approximations are in use for
both resonating and non-resonating processes, dressed by initial-state
QED corrections.  A status report can be found in ref \cite{ww}. 
QED corrections with soft photon
exponentiation to unstable $W$ pair production are implemented 
in Monte Carlo generators \cite{jadachetal}.

One of the specific problems in the theoretical description of the
production process for  $W$ bosons is the presence of the
width term in the $W$ propagator, which violates gauge invariance,       
yielding gauge-dependent amplitudes. As a solution, it has been
proposed \cite{argyres} to take into account also the imaginary
part in the $WW\gamma$ vertex from the light fermion triangle
loops (see also \cite{gabriel}).
This prescription is in accordance with gauge invariance
and cures the Ward identities between 2- and
3-point functions involving $W^{\pm}$ and $\gamma$.
This scheme can be extended to
incorporating the whole set of fermion loop contributions at
the one-loop level in the double- and single-resonating   
processes \cite{fermionscheme}.

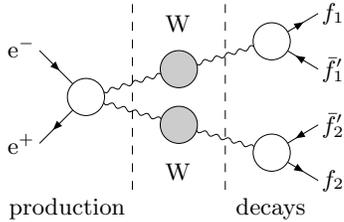
\begin{figure}
  \begin{center}
  \unitlength .7pt\small\SetScale{0.7}
  \begin{picture}(200,130)(0,-10)
    \ArrowLine(43,58)(25,40)        \Text(8,40)[lc]{\Pep}
    \ArrowLine(25,90)(43,72)        \Text(8,92)[lc]{\Pem}
    \Photon(50,65)(150,95){1}{16}   \Text(100,105)[]{\PW}  
    \Photon(50,65)(150,35){1}{16}   \Text(100,25)[]{\PW}
    \ArrowLine(155,98)(175,110)     \Text(190,110)[rc]{$f_1$}
    \ArrowLine(175,80)(155,92)      \Text(190,80)[rc]{$\bar{f}_1'$}
    \ArrowLine(175,50)(155,38)      \Text(190,50)[rc]{$\bar{f}_2'$}
    \ArrowLine(155,32)(175,20)      \Text(190,20)[rc]{$f_2$}
    \DashLine(75,115)(75,15){5}     \Text(40,5)[]{production}
    \DashLine(125,115)(125,15){5}   \Text(150,5)[]{decays}
    \GCirc(50,65){10}{1}
    \GCirc(100,80){10}{0.8}
    \GCirc(100,50){10}{0.8}
    \GCirc(150,95){10}{1}
    \GCirc(150,35){10}{1}
  \end{picture}
  \end{center}
  \caption[]{The generic structure of the factorizable
             W-pair contributions. The shaded circles indicate the 
             Breit--Wigner resonances.}
  \label{double-pole}
\end{figure}

\begin{figure}
  \begin{center}
  \unitlength .7pt\small\SetScale{0.7}
  \begin{picture}(120,100)(0,0)
  \ArrowLine(30,50)( 5, 95)
  \ArrowLine( 5, 5)(30, 50)
  \Photon(30,50)(90,80){2}{6}
  \Photon(30,50)(90,20){2}{6}
  \GCirc(30,50){10}{0}
  \Vertex(90,80){1.2}
  \Vertex(90,20){1.2}
  \ArrowLine(90,80)(120, 95)
  \ArrowLine(120,65)(105,72.5)
  \ArrowLine(105,72.5)(90,80)
  \Vertex(105,72.5){1.2}
  \ArrowLine(120, 5)( 90,20)
  \ArrowLine( 90,20)(105,27.5)
  \ArrowLine(105,27.5)(120,35)
  \Vertex(105,27.5){1.2}
  \Photon(105,27.5)(105,72.5){2}{4.5}
  \put(92,47){$\gamma$}
  \put(55,73){\PW}
  \put(55,16){\PW}
  \end{picture}
  \quad
  \begin{picture}(120,100)(0,0)
  \ArrowLine(30,50)( 5, 95)
  \ArrowLine( 5, 5)(30, 50)
  \Photon(30,50)(90,80){2}{6}
  \Photon(30,50)(90,20){2}{6}
  \Vertex(60,35){1.2}
  \GCirc(30,50){10}{0}
  \Vertex(90,80){1.2}
  \Vertex(90,20){1.2}
  \ArrowLine(90,80)(120, 95)
  \ArrowLine(120,65)(105,72.5)
  \ArrowLine(105,72.5)(90,80)
  \Vertex(105,72.5){1.2}
  \ArrowLine(120, 5)(90,20)
  \ArrowLine(90,20)(120,35)
  \Photon(60,35)(105,72.5){2}{5}
  \put(87,46){$\gamma$}
  \put(63,12){\PW}
  \put(40,24){\PW}
  \put(55,73){\PW}
  \end{picture}
  \quad
  \begin{picture}(160,100)(0,0)
  \ArrowLine(30,50)( 5, 95)
  \ArrowLine( 5, 5)(30, 50)
  \Photon(30,50)(90,80){-2}{6}
  \Photon(30,50)(90,20){2}{6}
  \Vertex(60,65){1.2}
  \GCirc(30,50){10}{0}
  \Vertex(90,80){1.2}
  \Vertex(90,20){1.2}
  \ArrowLine(90,80)(120, 95)
  \ArrowLine(120,65)(105,72.5)
  \ArrowLine(105,72.5)(90,80)
  \Vertex(105,27.5){1.2}
  \ArrowLine(120, 5)(90,20)
  \ArrowLine(105,27.5)(120,35)
  \ArrowLine(90,20)(105,27.5)
  \Photon(60,65)(105,27.5){-2}{5}
  \put(84,55){$\gamma$}
  \put(63,78){\PW}
  \put(40,68){\PW}
  \put(55,16){\PW}
  \end{picture}
%
  \\[2ex]
  \unitlength .7pt\small\SetScale{0.7}
  \begin{picture}(240,100)(0,0)
  \ArrowLine(30,50)( 5, 95)
  \ArrowLine( 5, 5)(30, 50)
  \Photon(30,50)(90,80){2}{6}
  \Photon(30,50)(90,20){2}{6}
  \GCirc(30,50){10}{0}
  \Vertex(90,80){1.2}
  \Vertex(90,20){1.2}
  \ArrowLine(90,80)(120, 95)
  \ArrowLine(120,65)(105,72.5)
  \ArrowLine(105,72.5)(90,80)
  \ArrowLine(120, 5)( 90,20)
  \ArrowLine( 90,20)(120,35)
  \Vertex(105,72.5){1.2}
  \PhotonArc(120,65)(15,150,270){2}{3}
  \put(55,73){\PW}
  \put(55,16){\PW}
  \put(99,47){$\gamma$}
  \DashLine(120,0)(120,100){6}
  \PhotonArc(120,35)(15,-30,90){2}{3}
  \Vertex(135,27.5){1.2}
  \ArrowLine(150,80)(120,95)
  \ArrowLine(120,65)(150,80)
  \ArrowLine(120, 5)(150,20)
  \ArrowLine(150,20)(135,27.5)
  \ArrowLine(135,27.5)(120,35)
  \Vertex(150,80){1.2}
  \Vertex(150,20){1.2}
  \Photon(210,50)(150,80){2}{6}
  \Photon(210,50)(150,20){2}{6}
  \ArrowLine(210,50)(235,95)
  \ArrowLine(235, 5)(210,50)
  \GCirc(210,50){10}{0}
  \put(177,73){\PW}
  \put(177,16){\PW}
  \end{picture}
  \end{center}
  \caption[]{Examples of virtual (top) and real (bottom) non-factorizable 
             corrections to \PW-pair production.}
  \label{fi:nf}
\end{figure}
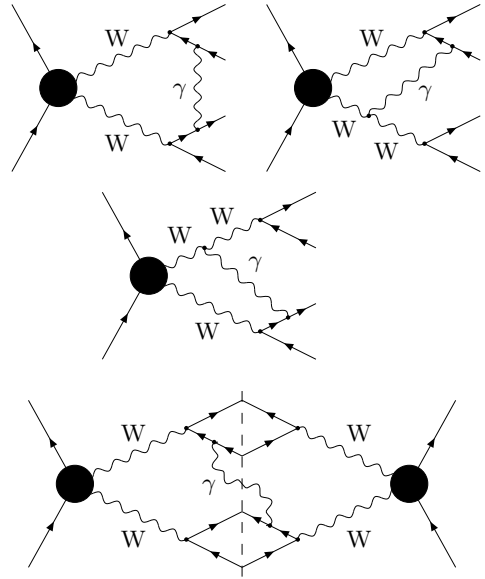

\begin{figure}
  \centerline{ 
  \setlength{\unitlength}{1cm}
  \begin{picture}(8,7)
  \put(-.8,-3.5){\includegraphics{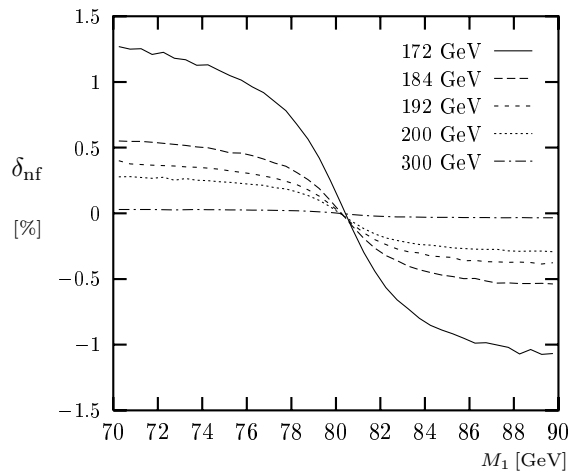}}
  \put(-0.3,3.7){\makebox(1,1)[c]{$\delta_\nf$}}
  \put(-0.3,2.9){\makebox(1,1)[c]{\scriptsize [\%]}}
  \put(6.3,-0.2){\makebox(1,1)[cc]{{$\scriptstyle{M_1}\,
                                   [\mbox{\scriptsize GeV}]$}}}
  \end{picture}
  }
  \caption[]{Relative non-factorizable corrections  
    to the single-invariant-mass distribution in the process
    $e^+e^-\ra WW \ra e^+\nu_e e^- \bar{\nu}_e$.
    From ref \protect{\cite{nonfactor2}}. } 
\label{energyplot}
\end{figure}

The  systematic treatment of the 
complete radiative corrections is  a task of enormous 
complexity. A reasonable simplification is given in terms
of the double-pole approximation with two resonating
$W$ bosons, the accuracy of which is estimated to be 
of order 0.1\% if one is not too close to the $WW$ threshold
\cite{beenakkerdenner}.
The `factorizable corrections',
displayed in Figure \ref{double-pole}, 
can be attributed either to the production of the gauge boson
pair or to the subsequent decays.
In the other class of `non-factorizable corrections' the diagrams
cannot be separated into a production process 
and  decay processes
(Figure \ref{fi:nf}). There are two recent independent 
calculations of the non-factorizable corrections in the
double-pole approximation 
\cite{nonfactor1,nonfactor2}, with very good  
agreement.
Their effect on the invariant mass distribution of one of the
decaying $W$'s is below 1\% for energies above 180 GeV.
An example is shown in Figure \ref{energyplot}, where 
the single invariant-mass 
distribution ${\rm d}\sigma/{\rm d} M_1$
is displayed for the process 
$e^+e^-\ra WW \ra e^+\nu_e e^- \bar{\nu}_e$. The signatures are
very similar also for other decay channels.
In the inclusive cross sections, the non-factorizable terms are
zero in the double-pole approximation \cite{yakovlev}.

In the case of 4-quark final states, diagrams similar to those 
in Figure \ref{fi:nf} arise, with the photon between two
fermions replaced by a gluon line.
Such typical non-factorizable QCD corrections are only
formally analogous to the QED ones: in the soft-gluon limit,
which is required to maintain the double-pole structure of
the amplitude, the strong interaction becomes non-perturbative
and can thus not be dealt with in terms of Feynman diagrams.
This `colour reconnection' leads to a distortion of the individual
hadronic systems from separated $W$ decays and can at present
be treated only with the help of hadronization models. It 
yields the dominant systematic error in the $W$ mass
reconstruction from 4-jet final states and is 
mainly responsible 
for the limited accuracy of about 40 MeV in the $W$-mass 
measurement at LEP.

Production of single-$W$ resonances occurs as a Drell--Yan process
 $ q q' \ra W \ra \ell^+ \nu_{\ell}$ in hadron collisions.
Run II of the upgraded Tevatron will provide precision 
measurements of $M_W$, comparable to that at LEP or even more 
accurate. For this purpose the inclusion of the
complete set of one-loop
electroweak corrections to the resonating Drell--Yan process 
\cite{baur,wackeroth} becomes necessary.
The electroweak radiative corrections to the $W$ propagator
around the resonance have also been studied in \cite{passera}.

\section{The Higgs sector}
The minimal model with a single scalar doublet is the simplest way
to implement the electroweak symmetry breaking. The experimental
result that the $\rho$-parameter is very close to unity is a
natural feature of models with doublets and singlets.
In the standard model, the mass $M_H$ of the Higgs boson
appears as the only additional parameter beyond the vector boson
and fermion masses. $M_H$ cannot be predicted but has to be taken from
experiment. The present lower limit (95\% C.L.) from the search at
LEP \cite{treille} is 89 GeV.
Indirect determinations of $M_H$ from precision data have
already been discussed in section 3. The indirect mass bounds
react  sensitively to small changes in the input data,
which is a consequence of the logarithmic dependence of 
electroweak precision observables. 
As a general feature,
it appears that the data prefer a light Higgs boson.

\begin{figure}[htb]
\centerline{
\epsfig{figure=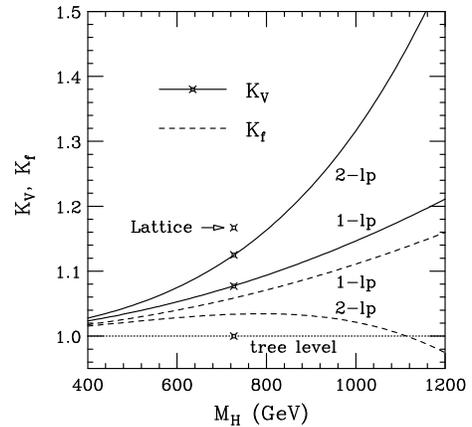,height=6cm,angle=90}}
\caption{Correction factors for the Higgs decay
         widths $H\ra VV\; (V=W,Z)$ and 
         $H \ra f \bar{f}$ in 1- and 2-loop order 
         (from ref {\protect\cite{riess}}) }  
\label{kfactors}
\end{figure}

\medskip
There are also theoretical constraints on the Higgs mass
from vacuum stability and absence of a Landau pole 
\cite{lindner,higgsbounds,hambye},
and from lattice calculations \cite{lattice}. 
Explicit perturbative calculations
of the decay width for $H\ra W^+W^-,ZZ$  in the large-$M_H$ limit
in 2-loop order \cite{ghinculov} have  shown that the 2-loop
contribution exceeds the 1-loop term in size (same sign) for
 $M_H > 930$ GeV (Figure \ref{kfactors}).
This result  is confirmed by the calculation of the next-to-leading
order
correction  in the $1/N$ expansion, where the Higgs sector is treated
as an $O(N)$ symmetric 
$\sigma$-model \cite{binoth}. 
A similar increase of the 2-loop perturbative contribution 
with $M_H$
is observed for the fermionic 
decay \cite{Hff}  $H\ra f \bar{f}$, but with opposite sign 
leading to a cancellation of the one-loop correction 
for $M_H\simeq 1100$ GeV (Figure \ref{kfactors}).
The requirement of applicability of
perturbation theory therefore puts a stringent upper limit on the
Higgs mass. The indirect Higgs mass bounds obtained from the
precision analysis show, however, that the Higgs boson is well below
the mass range where the Higgs sector becomes non-perturbative.
The lattice result \cite{decaylattice} for the bosonic Higgs decay
in Figure \ref{kfactors} for $M_H=727$ GeV is not far from
the perturbative 2-loop result. 
The difference may at least partially be interpreted as missing 
higher-order terms.


\begin{figure}[htb]
\centerline{
\epsfig{figure=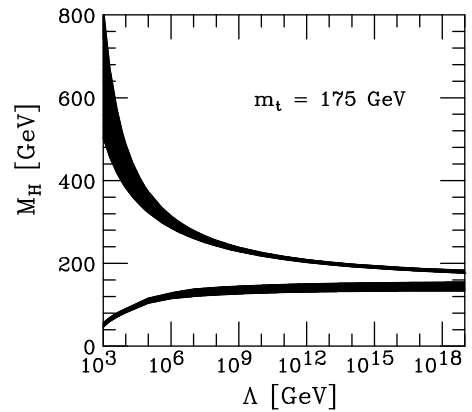,height=6cm,angle=90}}
\caption{Theoretical limits on the Higgs boson mass from
         the absence of a Landau pole and from vacuum stability
         (from ref {\protect\cite{hambye}}) }  
\label{higgslimits}
\end{figure}  

\smallskip
The behaviour of the quartic Higgs self-coupling $\lambda$,
as a function of a rising energy scale $\mu$, follows
from the renormalization group equation with the $\beta$-function
dominated by $\lambda$ and
the top quark Yukawa coupling $g_t$ contributions:
\beq
 \beta_{\lambda} = 24\, \lambda^2 + 12\, \lambda\,g_t^2
                    - 6 \, g_t^4 + \cdots
\eeq
In order to avoid unphysical negative quartic couplings
from the negative top quark contribution,
a lower
bound on the Higgs mass is derived. 
The requirement that the Higgs coupling remains finite and
positive up to a scale $\Lambda$ yields constraints
on the Higgs mass $M_H$, which have been evaluated at the 
2-loop level \cite{higgsbounds,hambye}.  
These bounds on $M_H$ are shown in Figure \ref{higgslimits}  
as a function of the cut-off scale $\Lambda$ up to which the
standard Higgs sector can be extrapolated,
for $m_t = 175$ GeV and $\al_s(M_Z) = 0.118$. The allowed 
region is the area between the lower and the upper curves.
The bands indicate the theoretical uncertainties associated
with the solution of the renormalization group equations
\cite{hambye}. It is interesting to note that the
indirect determination of the Higgs mass range from
electroweak precision data via radiative corrections 
is compatible with a value of $M_H$ where $\Lambda$ can
extend up to the Planck scale.

\section{The standard model at lower energies} 

\subsection{The decay $B \ra X_s \gamma$}

The rare radiative decay processes  $B \ra X_s \gamma$ 
are mediated by loop diagrams and hence represent sensitive
probes of the \sm as well as of extensions 
such as 2-Higgs doublet models 
or supersymmetric models.
In the \sm the next-to-leading QCD calculation 
for the total branching ratio has been completed
\cite{bsgammaQCD}, which together with the electroweak corrections
\cite{bsgammaEW} yields as the present best \sm prediction
(for a recent review see \cite{neubert}):
\beq
\label{bsgtheo}
   {\rm B}(B\ra X_s \g)_{\rm theor} = (3.29 \pm 0.33) \cdot 10^{-4} \, .
\eeq
From the experimental side, the CLEO Collaboration has
reported a new result: \cite{CLEO}
${\rm B}(B\ra X_s \g) = (3.15 \pm 0.35 \pm 0.32\pm 0.26) \cdot 10^{-4}$; 
the corresponding
result by ALEPH \cite{ALEPH} from $B$ mesons produced at the
$Z$ resonance is 
${\rm B}(B\ra X_s \g) = (3.11 \pm 0.80\pm 0.72) \cdot 10^{-4}$.
The experimental results are very close to each other and
agree remarkably well with the \sm prediction 
(\ref{bsgtheo}).
This further confirmation of the \sm
simultaneously puts stringent limits to
potential New Physics beyond the standard model
\cite{newphys}.

\subsection{Muon anomalous magnetic moment}
The anomalous magnetic moment of the muon,
\beq
   a_{\mu} = \frac{g_{\mu}-2}{2}
\eeq
provides a precision test of the standard model at low energies.
Within the present experimental accuracy \cite{PDG98} of
$\Delta\amu = 840\cdot 10^{-11}$, theory and experiment are in best
agreement, but the electroweak loop corrections are still hidden
in the noise. The new experiment E 821 at the Brookhaven National
Laboratory \cite{brookhaven}
is designed  to reduce
the experimental error down to $40\pm 10^{-11}$ and hence will
become sensitive to the electroweak loop contribution.

\begin{figure}[htb]
\centerline{
\epsfig{figure=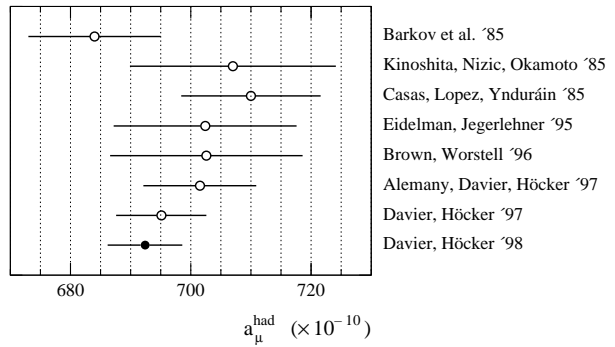,height=4.8cm}}
\caption{Various determinations of $a_{\mu}^{\rm had}$ 
         (from ref {\protect\cite{hoecker}}) }  
\label{amu}
\end{figure}

For this reason the standard model prediction has to be known 
with at least comparable precision. 
Recent theoretical work in this direction 
has provided the electroweak
2-loop terms \cite{ew22,ew23} with 3-loop 
leading-logarithmic contributions \cite{degrassigiudice}
and updated the contribution from the hadronic
photonic vacuum polarization
\cite{eidelman,dub,alemany,davier}, which is visualized in
Figure \ref{amu}.
The lowest data point with the smallest error \cite{davier}
is obtained with the help of the theory-driven QCD analysis,
which has been applied also to $\dal_{\rm had}(M_Z)$
(see section 2.1).


The         
main sources of the theoretical error at present are the hadronic
vacuum polarization and the light-by-light scattering mediated by
quarks, as part of the 3-loop hadronic contribution
\cite{sanda,hayakawa,bijnens}.
Table 3 contains  the breakdown of $\amu$. The hadronic part
is supplemented by the higher-order $\al^3$
vacuum polarization effects \cite{had3} 
(included in the numerical value), but it does not include the
light-by-light scattering contribution, which is listed separately
in the table.
 
\begin{table}[htbp]
\caption[]
{Contributions $\Delta\amu$ to the muonic anomalous magnetic moment
and their theoretical uncertainties, in units of $10^{-11}$.  }
\vspace{0.5cm}
\begin{center}
\begin{tabular}{@{}| l r r |}
\hline 
Source & $\Delta\amu$  & Error \\
\hline
 &  &  \\[-0.2cm]
QED \cite{qed} & 116584706 & 2  \\
Hadronic \cite{eidelman,had3}  & 6916  & 153 \\
Hadronic \cite{davier}  & 6816  &  62  \\
EW, 1-loop \cite{ew1} & 195 &     \\
EW, 2-loop \cite{ew23}  & $-44$   &  4  \\
Light-by-light \cite{hayakawa} & $-79$ &  15 \\
Light-by-light \cite{bijnens} & $-92$ &  32 \\[0.2cm]
\hline
Exp.\ (future)  &   & 40  \\
\hline
\end{tabular}
\end{center}
\label{amutab}
\end{table}

The 2-loop electroweak contribution is as large 
as the expected experimental error.
The dominating theo\-retical uncertainty at present is 
still the error in
the hadronic vacuum polarization. The previous discrepancy
in the contribution involving
light-by-light scattering  has been removed, with the consequence
that this term can now be considered as established with an 
acceptable uncertainty.

\section{Beyond the standard model}

\subsection{Conceptual problems}
The comparison of the theoretical predictions with
experimental data has confirmed the validity of the \sm in an 
impressive way: 

-- the description of the data is nearly
 perfect, with no significant indication for deviations;

-- the quantum effects of the \sm have been established at the
level of several $\sigma$; 

-- direct and indirect determinations 
of the top quark mass are compatible with each other;

-- the Higgs boson mass is meanwhile also being constrained
within the perturbative mass regime
with the possibility that the \sm may be extrapolated up to energies
around the Planck scale.
 
 \smallskip \noi 
In spite of this success, the conceptual situation with
the \sm is unsatisfactory for quite a few deficiencies:

-- the smallness of the electroweak scale 
$v \sim 246\, \gv << M_{\rm Pl}$ (the `hierarchy problem'); 

-- the large number of free parameters 
(gauge couplings, vacuum expectation value, $M_H$, fermion
masses, CKM matrix elements), which are not predicted but
have to be taken from experiments; 

-- the pattern that occurs in the arrangement of the fermion 
masses;

-- the missing way to connect to gravity.

\noi
It is a curiosity of the \sm that these questions
will persist even after the Higgs boson will
have been be discovered.

\subsection{Massive neutrinos}
Besides the long-standing list of conceptual theoretical problems,
a new perspective arises through the recent experimental results
by Super-Kamiokande \cite{superka}
on the atmospheric neutrino anomaly, which can most easily be
explained by  oscillations between different $\nu$ species,
associated with neutrino masses different from zero. 
In the strict-minimal model, neutrinos are massless and right-handed
neutrino components are absent.
The evidence for massive neutrinos requires a modification of
the minimal model in order to accommodate neutrinos with mass.
The straightforward way to introduce mass terms is the
augmentation of the fermion sector by right-handed partners
$\nu_{\rm R}$;  together with the familiar $\nu_{\rm L}$ these allow
the presence of Dirac mass terms 
$\sim m_{\nu} \bar{\nu} \nu$ with $ \nu = \nu_{\rm L} + \nu_{\rm R}$ and
$m_{\nu}$ as additional mass parameters, without altering  
the global architecture of the standard model and without
spoiling the successful description of all the other 
electroweak phenomena.  
What appears  unsatisfactory is the unexplained smallness
of the neutrino Dirac masses, as enforced by the
empirical situation. A commonly accepted elegant solution  
is given by the seasaw-mechanism where a 
lepton-number-violating 
Majorana mass term $\sim M$ is introduced. 
Together with the Dirac mass, which is of the order of the
usual charged lepton masses, a very light and a very heavy
$\nu$ component appear with the light one almost entirely
left-handed, when $M$ is of the order of the GUT scale.
Candidates for specific models are
Grand Unification scenarios 
such as the SO(10)-GUT, where $\nu_{\rm R}$
fits into the same 16-dimensional representation as the other
fermions of a family.
Hence, the appearance of small neutrino masses
points towards a new high-mass scale 
beyond the minimal model, which may be 
associated with the concept of further unification
of the fundamental forces.

\subsection{The minimal supersymmetric standard model (MSSM)} 
Among the extensions of the standard model, the MSSM is
the theoretically favoured scenario as 
the most predictive framework beyond the standard model.
A definite prediction of the MSSM
is the existence of a light Higgs boson
with mass below $\sim 135$ GeV \cite{hmass}. The detection of a
light Higgs boson at LEP could be a significant hint for
supersymmetry.

The structure of the MSSM as a renormalizable quantum field theory
allows a similarly complete calculation of
the electroweak precision observables
as in the standard model in terms of one Higgs mass
(usually taken as the $CP$-odd `pseudoscalar' mass
$M_A$) and $\tan\beta= v_2/v_1$,
together with the set of
SUSY soft-breaking parameters fixing the chargino/neutralino and
scalar fermion sectors.
It has been known for quite some time
\cite{higgs}
that light non-standard
Higgs bosons as well as light stop and charginos
predict larger values for the ratio $R_b$ \cite{susy1,susy3}.
Complete 1-loop calculations are available for
$\Delta r$ \cite{susydelr} and for the $Z$ boson observables
\cite{susy3}.

\smallskip
A possible mass splitting between
 $\tilde{b}_L$ and $\tilde{t}_L$  
yields a contribution to the $\rho$-parameter
of the same sign as the standard top term.
As a universal loop contribution, it enters the quantity $\Dr$ and the
$Z$ boson couplings
and is thus significantly constrained by the data on $M_W$ and the
leptonic widths. Recently the 2-loop $\al_s$ corrections have been
computed \cite{drosusy},
which can amount to 30\% of the 1-loop
$\Delta \rho_{\tilde{b}\tilde{t}}$.

\begin{figure}[htb]
\vspace{-2cm}
\centerline{
\mbox{\epsfxsize8.0cm\epsffile{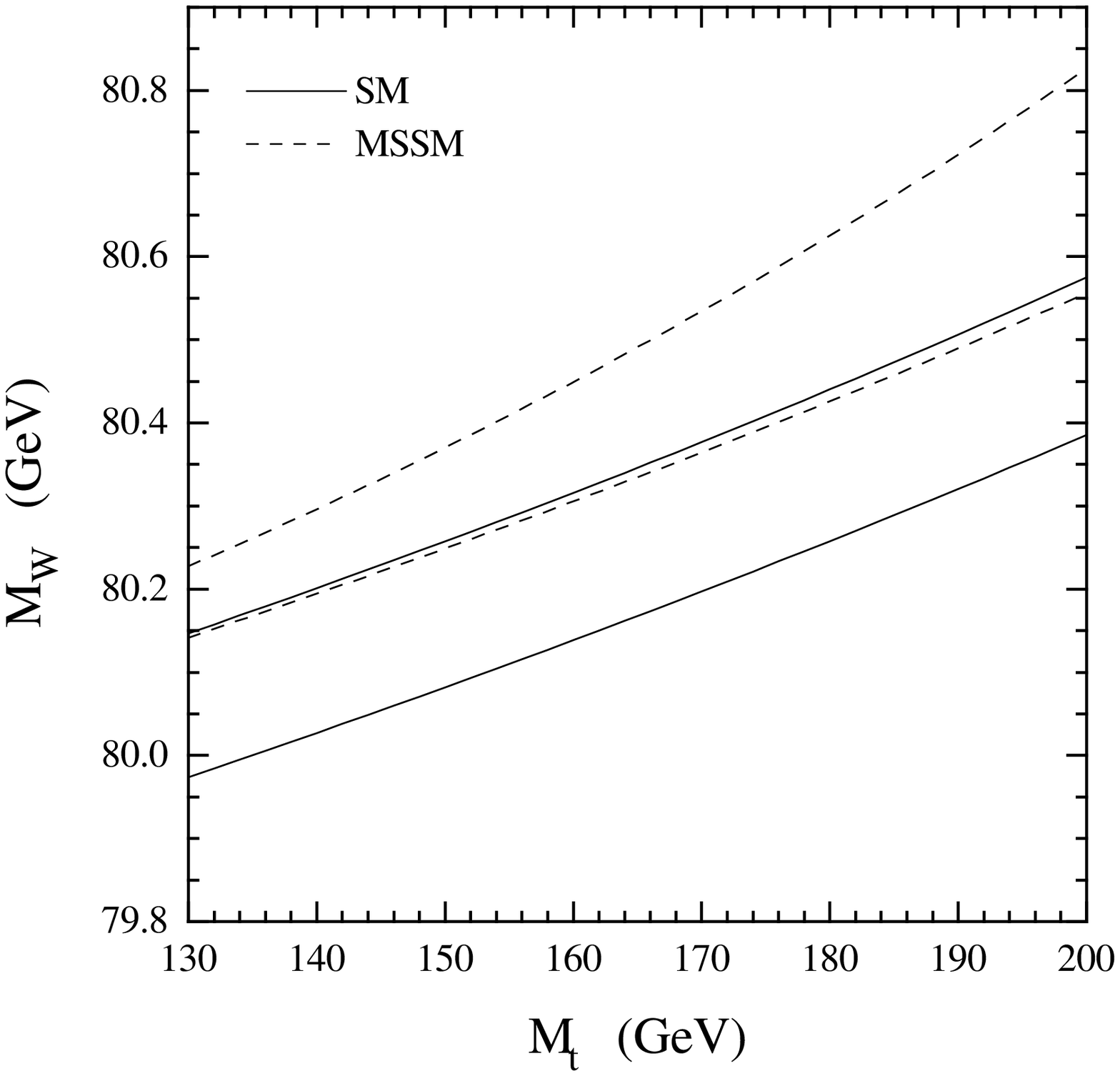}}} 
\vspace{-3cm}
\caption{The $W$ mass range in the standard model (-----) and the
         MSSM (- - -). Bounds are from the non-observation of Higgs
         bosons and SUSY particles at LEP2.} 
\label{susymw}
\end{figure}

\smallskip 
Figure \ref{susymw}
displays the range of predictions for $M_W$ in the minimal model
and in the MSSM. It is thereby assumed that no direct discovery has been
made at LEP~2. As can be seen, precise determinations of $M_W$ and $m_t$
can become decisive for the separation between the  models.
 
\setlength{\unitlength}{0.7mm}

\begin{figure}[htb]
\centerline{
\mbox{\epsfxsize6.0cm\epsffile{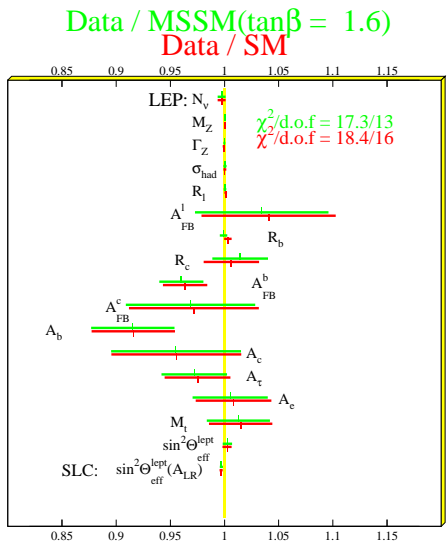}}} 
\vspace{-1cm}
\caption{Best fits in the SM and in the MSSM, normalized to 
         the data. Error bars are those from data.
         (Updated from ref {\protect\cite{deboer}. })  } 
\label{balken}
\end{figure}

\smallskip
As the standard model,
the MSSM yields a  good description of the precision data.
A global fit 
\cite{deboer}
to all electroweak precision data, including the top mass measurement,
shows that
 the $\chi^2$ of the fit is slightly better than in the
standard model;
 but, owing to the larger numbers of parameters, the
probability is about the same as for the 
standard model (Figure \ref{balken}).

\setlength{\unitlength}{0.7mm}

The virtual presence of SUSY particles in the precision
observables can be exploited also in the other way of
constraining the allowed range of the MSSM parameters. 
Since the quality of the \sm description can be achieved
only for those parameter sets where the \sm with a light
Higgs boson is approximated, deviations from this
scenario result in a rapid decrease of the fit quality.
An analysis of the precision data in this spirit can be
found in ref \cite{pierce}.

\section{Conclusions}
The experimental data for tests of the standard model 
have achieved an impressive accuracy.
In the meantime, many 
theoretical contributions have become available to improve and 
stabilize the \sm predictions and to
reach a theoretical
accuracy clearly better than 0.1\%.

The overall agreement between
theory and experiment for the entire  set of the
precision observables is remarkable and instructively confirms the
validity of the standard model. Fluctuations of data around the
predictions are within two standard deviations, with no compelling
evidence for deviations.
Direct and indirect determinations of the top mass are compatible, and 
a light Higgs boson is clearly favoured by the analysis of precision
data in the \sm context, which is far below the mass range where
the standard Higgs sector becomes non-perturbative.

As a consequence of the high quality performance of the standard model,
any kind of New Physics can only provoke small effects, at most of the 
size that is set by the radiative corrections.
The MSSM, mainly theoretically advocated,
is competitive to the standard model in describing the data
with about the same quality in global fits.
Since the MSSM predicts the existence of a light Higgs boson,
the detection of a Higgs at LEP could be an indication
of supersymmetry. The \sm can also accommodate such a light
Higgs, but with the consequence that its validity cannot 
be extrapolated to energies much higher than the TeV scale.

\section*{Acknowledgements}

I want to thank
D. Bardin, G. Degrassi, A. Denner, S.~Ditt\-maier, 
J.~Erler, P.~Gambino,
F.~Jegerlehner, J.~K\"orner, J.~K\"uhn, M.~Neubert,
G.~Passarino,
B.~Ward, G.~Weiglein
 for helpful discussions and support,
W.~de Boer and U.~Schwickerath for updated fits in the MSSM, and
A.~Blondel, R.~Clare, M.~Gr\"unewald, D.~Karlen for providing me
with valuable information  on the experimental data.

\end{document}

%% file: titlepage.tex
\thispagestyle{empty}
\setcounter{page}{0}
\def\thefootnote{\fnsymbol{footnote}}

{\textwidth 17cm

\twocolumn[

\begin{flushright}
CERN-TH/98-358 \\
KA--TP--18--1998\\
hep-ph/9811313 \\
\end{flushright}

\vspace{2cm}

\begin{center}

{\large\sc {\bf STANDARD MODEL THEORY}}

\vspace{2cm}

{\sc W. Hollik }

\vspace*{1cm}

     Theoretical Physics Division, CERN \\
     CH-1211 Geneva 23, Switzerland
     
\vspace*{0.4cm}
  
      and 

\vspace*{0.4cm}

     Institut f\"ur Theoretische Physik, Universit\"at Karlsruhe \\
     D-76128 Karlsruhe, Germany


\end{center}

\vspace*{2cm}

\begin{abstract}
In this conference report a summary is given on the theoretical 
work that has contributed to provide accurate 
theoretical predictions for testing the \sm in present and future 
experiments. Precision calculations for the  vector boson masses,
for the $Z$ resonance, $W$ pair production, and for the $g-2$ of the muon
are reviewed and the theoretical situation for the Higgs sector
is summarized.
 The status of the \sm is discussed 
in the light of the recent 
high and low energy data.
New Physics beyond the \sm is briefly addressed
as well, with special
emphasis on the minimal supersymmetric standard model.
\end{abstract}

\vspace*{4cm}

\begin{center}

           Plenary talk at the \\
           XXIX International Conference on
           High Energy Physics \\ Vancouver, Canada,
           23 - 29 July 1998

\end{center}

]

}
\def\thefootnote{\arabic{footnote}}
\setcounter{footnote}{0}

\newpage

%% file: sm1.bbl
\section*{References}

\begin{thebibliography}{99}
\bibitem{lep}
The LEP Collaborations ALEPH, DELPHI, L3, OPAL, the LEP Electroweak
Working Group and the SLD Heavy Flavor Working Group,
CERN-PPE/97-154 (1997); \\ M. Gr\"unewald, talk at ICHEP98 (PS1), these
          proceedings
\bibitem{karlen}
                 D. Karlen, plenary talk at ICHEP98,
                 these proceedings
\bibitem{wmass}  UA2 Collaboration, J. Alitti et al.,
                 \PLB {\bf 276}, 354 (1992); \\
                 CDF Collaboration, F. Abe et al.,
                 \PRL {\bf 65}, 2243 (1990); 
                 \PRD {\bf 43}, 2070 (1991); 
                 \PRL {\bf 75}, 11 (1995);
                 \PRD {\bf 52}, 4784 (1995); \\
                 D0 Collaboration, B. Abbott et al.,
                 \PRL {\bf 80}, 3008 (1998); \\  
                 P. Derwent, A. Kotwal, talks at ICHEP98 (PS1), 
                 these proceedings
\bibitem{top}
CDF Collaboration, F. Abe et al., \PRL {\bf 74}, 2626 (1995); \\
D0 Collaboration, S. Abachi et al., \PRL {\bf 74}, 2632 (1995)
\bibitem{partridge} R. Partridge, plenary talk at ICHEP98, 
                    these proceedings
\bibitem{thooft}
G. 't Hooft, \NPB {\bf 33}, 173 (1971) and B
              {\bf 35}, 167 (1971)
\bibitem{pav}
G. Passarino, M. Veltman, \NPB {\bf 160}, 151 (1979)
\bibitem{con}
M. Consoli, \NPB {\bf 160}, 208 (1979)
\bibitem{sirmar}
A. Sirlin, \PRD {\bf 22}, 971 (1980); \\
W. J. Marciano, A. Sirlin, \PRD {\bf 22}, 2695 (1980); \\
A. Sirlin, W. J. Marciano, \NPB {\bf 189}, 442 (1981)
\bibitem{dubna}
D.Yu.\ Bardin, P.Ch.\ Christova, O.M. Fedorenko,
\NPB {\bf 175}, 435 (1980) and B {\bf 197}, 1 (1982); \\
D.Yu.\ Bardin, M.S. Bilenky, G.V. Mithselmakher, T. Riemann,
M. Sachwitz,
\ZPC {\bf 44}, 493 (1989)
\bibitem{fjeg}
J. Fleischer, F. Jegerlehner, \PRD {\bf 23}, 2001 (1981)
\bibitem{aoki}
K.I. Aoki, Z. Hioki, R. Kawabe, M. Konuma,
    \\ T. Muta,
{\it Suppl.\ Prog.\ Theor.\ Phys.\ } {\bf 73},1 (1982); \\
Z. Hioki, \PRL {\bf 65}, 683 (1990), 
  E: {\em ibidem} {\bf 65}, 1692 (1990);
  \ZPC {\bf 49}, 287 (1991)
\bibitem{maiani}
M. Consoli, S. LoPresti, L. Maiani,
 \NPB {\bf 223}, 474 (1983)
\bibitem{dz}
D.Yu.\ Bardin, M.S. Bilenky, G.V. Mithselmakher, T. Riemann,
M. Sachwitz,
\ZPC {\bf 44}, 493 (1989)
\bibitem{bhs} M. B\"ohm, W. Hollik, H. Spiesberger,
{\em Fortschr.\ Phys.\ } {\bf 34}, 687 (1986)
\bibitem{hollik}
W. Hollik, {\it Fortschr.\ Phys.\ } {\bf 38}, 165 (1990)
\bibitem{LEP}
M. Consoli, W. Hollik, F. Jegerlehner, in:
{\it
$Z$ Physics at LEP 1},
  eds.\ G. Altarelli, R. Kleiss
and C. Verzegnassi,
 CERN 89-08   (1989)
\bibitem{pittau}
G. Passarino, R. Pittau, \PLB {\bf 228}, 89 (1989); \\
V.A. Novikov, L.B. Okun, M.I. Vysotsky, 
             \NPB {\bf 397}, 35 (1993)
\bibitem{pavelt}
G. Passarino, M. Veltman, \PLB {\bf 237}, 537 (1990)
\bibitem{msbar}
W.J. Marciano, A. Sirlin, \PRL {\bf 46}, 163 (1981); \\
A. Sirlin, \PLB {\bf 232}, 123 (1989)
\bibitem{msbar1}
G. Degrassi, S. Fanchiotti, A. Sirlin,
\NPB {\bf 351}, 49 (1991) 
\bibitem{msbar2}
G. Degrassi, A. Sirlin,
\NPB {\bf 352}, 342 (1991) 
\bibitem{green}
M. Veltman, \PLB {\bf 91}, 95 (1980); \\ 
 M. Green, M. Veltman, \NPB
{\bf 169}, 137 (1980), E: \NPB {\bf 175}, 547 (1980); \\
F. Antonelli, M. Consoli, G. Corbo, \PLB {\bf 91}, 90 (1980); \\
F. Antonelli, M. Consoli, G. Corbo, O. Pellegrino,
\NPB {\bf 183}, 195 (1981)
\bibitem{star}
D.C. Kennedy, B.W. Lynn, \NPB {\bf 322}, 1 (1989)
\bibitem{star1}
M. Kuroda, G. Moultaka, D. Schildknecht, \NPB {\bf 350}, 25 (1991)
\bibitem{yb95}  {\it Reports of
             the Working Group on Precision Calculations
             for the $Z$ Resonance}, CERN 95-03 (1995), eds.\
             D. Bardin, W. Hollik, G. Passarino
\bibitem{kallen} G. K\"all\'en, A. Sabry,
               {\em K. Dan.~Vidensk.~Selsk.~Mat.-Fys.~Medd.}
               {\bf 29} (1955) No.~17
\bibitem{steinhauser1} M. Steinhauser, \PLB {\bf 429}, 158 (1998)
\bibitem{eidelman} S. Eidelman, F. Jegerlehner,
                   \ZPC {\bf 67}, 585 (1995)
\bibitem{burkhardt} H. Burkhardt, B. Pietrzyk,
                    \PLB {\bf 356}, 398 (1995)
\bibitem{swartz} M.L. Swartz, \PRD {\bf 53}, 5268 (1996)
\bibitem{alemany}
                 R. Alemany, M. Davier, A. H\"ocker,
                 {\it Eur.~Phys.~J.} C {\bf 2} (1998) 123
\bibitem{davier} M. Davier, A. H\"ocker, \PLB {\bf 419}, 419 (1998); 
                 hep-ph/9801361; hep-ph/9805470
\bibitem{hoecker} A. H\"ocker, talk at ICHEP98 (PS1), 
                   these proceedings  
\bibitem{kuehnsteinh} J.H. K\"uhn, M. Steinhauser, hep-ph/9802241
\bibitem{mainz} S. Groote, J. K\"orner, K. Schilcher, N.F. Nasrallah, 
                 hep-ph/9802374
\bibitem{erler} J. Erler, hep-ph/9803453
\bibitem{mqcd} A.H. Hoang, J.H. K\"uhn, T. Teubner,
               \NPB {\bf 452}, 173 (1995); \\
               K.G. Chetyrkin, J.H. K\"uhn, M. Steinhauser,
               \PLB {\bf 371}, 93 (1996); 
               \NPB {\bf 482}, 213 (1996); B {\bf 505}, 40 (1997); \\
               K.G.~Chetyrkin, R.~Harlander, J.H.~K\"uhn, M.~Steinhauser,
               \NPB {\bf 503}, 339 (1997)
\bibitem{martinzepp}
                 A.D. Martin, D. Zeppenfeld,
                 \PLB {\bf 345}, 558 (1995)
\bibitem{braaten} E. Braaten, S. Narison, A. Pich,
                  \NPB {\bf 373} (1992) 581
\bibitem{zhan} Z. Zhan, talk at ICHEP98 (PS1), these proceedings
\bibitem{fredjeger} F. Jegerlehner, IVth International Symposium on
                   Radiative Corrections, Barcelona, September 1998
                   (to appear in the proceedings, ed.~J. Sol\`a); \\
                   F. Jegerlehner, O.V. Tarasov, hep-ph/9809485
\bibitem{rho}
D. Ross, M. Veltman, \NPB {\bf 95}, 135 (1975)
\bibitem{rho1}
M. Veltman, \NPB  {\bf 123}, 89 (1977); \\
M.S. Chanowitz, M.A. Furman, I. Hinchliffe, \PLB {\bf 78}, 285 (1978)
\bibitem{bij} J.J. van der Bij, F. Hoogeveen, \NPB {\bf 283}, 477
               (1987)
\bibitem{barbieri}
R. Barbieri, M. Beccaria, P. Ciafaloni, G. Curci, A.~Vicere,
 \PLB {\bf 288}, 95 (1992); \NPB {\bf 409}, 105 (1993); \\
J. Fleischer, F. Jegerlehner, O.V. Tarasov, \PLB {\bf 319}, 249 (1993)
\bibitem{djouadi}
 A. Djouadi, C. Verzegnassi, \PLB {\bf 195}, 265 (1987)
\bibitem{tarasov} L. Avdeev, J. Fleischer, S. M. Mikhailov, O. Tarasov,
                  \PLB {\bf 336}, 560 (1994); 
                  E: \PLB {\bf 349}, 597 (1995); \\
                  K.G. Chetyrkin, J.H. K\"uhn, M. Steinhauser,
                  \PLB {\bf 351}, 331 (1995)
\bibitem{screening}
M. Veltman, {\em Acta Phys.\ Polon.\  } B {\bf 8}, 475 (1977)
\bibitem{bij1} J.J. van der Bij, M. Veltman, 
               \NPB {\bf 231}, 205 (1985)
\bibitem{muon}
R.E. Behrends, R.J. Finkelstein, A. Sirlin,
   {\em Phys.~Rev.\ } {\bf 101}, 866 (1956); \\
T. Kinoshita, A. Sirlin, 
   {\em Phys.~Rev.\ } {\bf 113}, 1652 (1959)
\bibitem{stuartvanritbergen} T. van Ritbergen, R. Stuart,
          hep-ph/9802341; hep-ph/9808283
\bibitem{PDG98}  Particle Data Group, 
              C. Caso et al., {\em Eur.~Phys.~J.\ C} {\bf 3}, 1 (1998)
\bibitem{kuehntalk} J.H. K\"uhn, talk at ICHEP98 (PS1), these
                   proceedings
\bibitem{marciano}
W.J. Marciano, \PRD {\bf 20}, 274 (1979)
\bibitem{chj}
M. Consoli, W. Hollik, F. Jegerlehner, \PLB {\bf 227}, 167 (1989)
\bibitem{qcd}
A. Djouadi, {\em Nuovo Cim.\ } A {\bf 100}, 357 (1988); \\
D. Yu.\ Bardin, A.V. Chizhov, Dubna preprint E2-89-525 (1989);\\
B.A. Kniehl, \NPB {\bf 347}, 86 (1990); \\ 
F. Halzen, B.A. Kniehl, \NPB {\bf 353}, 567 (1991) 567; \\
A. Djouadi, P. Gambino, \PRD {\bf 49}, 3499 (1994)
\bibitem{dispersion1}
B.A. Kniehl, J.H. K\"uhn, R.G. Stuart, \PLB {\bf 214}, 621 (1988);\\
B.A. Kniehl, A. Sirlin, \NPB {\bf 371}, 141 (1992); \\
                        \PRD {\bf 47}, 883 (1993);\\
 S. Fanchiotti, B.A. Kniehl, A. Sirlin, \PRD {\bf 48}, 307 (1993)
\bibitem{steinhauser}
K. Chetyrkin, J.H. K\"uhn, M. Steinhauser, \PRL {\bf 75}, 3394 (1995)
\bibitem{nonleading}
A. Sirlin, \PRD {\bf 29}, 89 (1984)
\bibitem{padova} G. Degrassi, P. Gambino, A. Vicini, 
                 \PLB {\bf 383}, 219 (1996); \\
  G. Degrassi, P. Gambino, A. Sirlin, \PLB {\bf 394}, 188 (1997); \\
  G. Degrassi, P. Gambino,  M.~Passera, A.~Sirlin,
                \PLB {\bf 418}, 209 (1998)
\bibitem{bauberger} S. Bauberger, G. Weiglein, 
             {\em Nucl.~Instrum.~Meth.~A} {\bf 389}, 318 (1997);
             \PLB {\bf 419}, 333 (1997)
\bibitem{weiglein} G. Weiglein, 
                   {\em Acta Phys.~Polon.\ B} {\bf 29}, 2735 (1998) 
\bibitem{achim1} W. Hollik, B. Krause, A. Stremplat, G. Weiglein,
                  to appear; \\
                A. Stremplat, Diploma Thesis (Karlsruhe 1998)
\bibitem{ewgr} D. Bardin et al., 
                {\it Reports of
             the Working Group on Precision Calculations
             for the $Z$ Resonance}, p.~7, 
             CERN 95-03 (1995), eds.\
             D.~Bardin, W.~Hollik, G.~Passarino;
              hep-ph/9709229
\bibitem{padova1}  G. Degrassi, P. Gambino, A. Sirlin, 
                 TUM-HEP-333/98 (to appear),
                 and  private communication
\bibitem{vertex}
A.A. Akhundov, D.Yu.\ Bardin, T. Riemann, \NPB {\bf 276}, 1 (1986);\\
W. Beenakker, W. Hollik, \ZPC {\bf 40}, 141 (1988);\\
J. Bernabeu, A. Pich, A. Santamaria, \PLB {\bf 200}, 569 (1988)
\bibitem{dhl} A. Denner, W. Hollik, B. Lampe,
              \ZPC {\bf 60}, 193 (1993)
\bibitem{qcdq}
K.G. Chetyrkin, A.L. Kataev, F.V. Tkachov, \PLB {\bf 85}, 277 (1979);\\
M. Dine, J. Sapirstein, \PRL {\bf 43}, 668 (1979);\\
W. Celmaster, R. Gonsalves, \PRL {\bf 44}, 560 (1980);\\
S.G. Gorishny, A.L. Kataev, S.A. Larin, \PLB {\bf 259}, 144 (1991);\\
L.R. Surguladze, M.A. Samuel, \PRL {\bf 66}, 560 (1991);\\
A. Kataev, \PLB {\bf 287}, 209 (1992)
\bibitem{qcdb}
  K.G. Chetyrkin, J.H. K\"uhn, \PLB {\bf 248}, 359 (1990) and
                                  B {\bf 406}, 102 (1997); \\
  K.G. Chetyrkin, J.H. K\"uhn, A. Kwiatkowski,
             \PLB {\bf 282}, 221 (1992);\\
  K.G. Chetyrkin, A. Kwiatkowski, \PLB {\bf 305}, 285 (1993) and
                                     B {\bf 319}, 307 (1993)
\bibitem{qcdb1}
    B.A. Kniehl, J.H. K\"uhn,  \PLB {\bf 224}, 229 (1990);
                               \NPB {\bf 329}, 547 (1990); \\
    K.G. Chetyrkin, J.H. K\"uhn, \PLB {\bf 307}, 127 (1993); \\  
    S. Larin, T. van Ritbergen, J.A.M. Vermaseren,
     \PLB {\bf 320}, 159 (1994); \\
    K.G. Chetyrkin, O.V. Tarasov, \PLB {\bf 327}, 114 (1994)
\bibitem{qcdq1}
         K.G.~Chetyrkin, J.H.~K\"uhn, A.~Kwiatkowski, in 
         {\it Reports of
             the Working Group on Precision Calculations
             for the $Z$ Resonance}, p.~175, 
             CERN 95-03 (1995), eds.\
             D.~Bardin, W.~Hollik, G.~Passarino; \\
             K.G. Chetyrkin, J.H. K\"uhn, A. Kwiatkowski, 
             {\em Phys.~Rep.\ } {\bf 277}, 189 (1996)
\bibitem{jeg} J. Fleischer, F.~Jegerlehner,
              P.~R\c{a}czka, O.V.~Tarasov,
              \PLB {\bf 293}, 437 (1992);\\
               G. Buchalla, A.J. Buras,
               \NPB {\bf 398}, 285 (1993);\\
              G. Degrassi, \NPB {\bf 407}, 271 (1993);\\
              K.G. Chetyrkin, A. Kwiatkowski, M. Steinhauser,
              {\em Mod.\ Phys. Lett.\ } A {\bf 8}, 2785 (1993)
\bibitem{log} A. Kwiatkowski, M. Steinhauser, \PLB {\bf 344}, 359 (1995);\\
              S. Peris, A. Santamaria, \NPB {\bf 445}, 252 (1995) 
\bibitem{harlander} R. Harlander, T. Seidensticker, M. Steinhauser,
                \PLB {\bf 426}, 125 (1998)
\bibitem{czarnecki}
A. Czarnecki, J.H. K\"uhn, \PRL {\bf 77}, 3955 (1996); E:
                           {\bf 80}, 893 (1998)
\bibitem{hoang} A. Hoang, J.H. K\"uhn, T. Teubner, \NPB {\bf 455}, 3
               (1995); {\bf 452}, 173 (1995)
\bibitem{berends} F.A. Berends et al.,
               in:
            {\it Z Physics at LEP 1}, CERN 89-08 (1989), eds.\
            G. Altarelli, R. Kleiss, C. Verzegnassi, Vol.\ I, p.\  89;\\
          W. Beenakker, F.A. Berends, S.C. van der Marck,
          \ZPC {\bf 46}, 687 (1990);\\
          G. Burgers, F.A. Berends, W.L. van Neerven,
              \NPB {\bf 297}, 429 (1988);
              E: {\bf 304}, 921 (1988);\\ 
             W. Beenakker, F.A. Berends, S.C. van der Marck,
             \ZPC {\bf 46}, 687 (1990)
\bibitem{montagna} S. Jadach, M. Skrzypek, B.F.L. Ward,
                   \PLB {\bf 257}, 173 (1991); \\
                  S. Jadach, M. Skrzypek, \ZPC {\bf 49}, 584 (1991);\\
                 G. Montagna, O. Nicrosini, F. Piccinini,
                 \PLB {\bf 406}, 243 (1997)
\bibitem{bhabha} S. Jadach et al., in
                  {\it Reports of
             the Working Group on Precision Calculations
             for the $Z$ Resonance}, CERN 95-03 (1995), p.\ 341, 
              eds.\
             D. Bardin, W.~Hollik, G.~Passarino; \\
             S. Jadach, O.Nicrosini, in
          {\em Physics at LEP 2}, CERN 96-01, Vol.\ 1,   
           eds.\ G. Altarelli,
          T. Sj\"ostrand, F.~Zwirner 
\bibitem{bennie} B.F.L. Ward, talk at ICHEP98 (PS1), these 
                proceedings; \\
         B.F.L. Ward, S. Jadach. M.Melles, S.A Yost,
         hep-ph/9811245
\bibitem{sld} SLD Collaboration, 
              S. Fahey, talk at ICHEP98 (PS1), these proceedings
\bibitem{gambinosirlin} P. Gambino, A. Sirlin,
                  \PRL {\bf 73}, 621 (1994)
\bibitem{schmelling} Yu.~Dokshitzer, plenary talk at ICHEP98,
                    these proceedings
\bibitem{neutrino}
G.L. Fogli, D. Haidt, \ZPC
{\bf 40}, 379 (1988);\\
CDHS Collaboration, H. Abramowicz et al.,
 \PRL {\bf 57}, 298 (1986);\\
A. Blondel et al., \ZPC {\bf 45}, 361 (1990);\\
CHARM Collaboration, J.V. Allaby et al.,
 \PLB {\bf 177}, 446 (1987);
\ZPC {\bf 36}, 611 (1987);\\
CHARM II Collaboration, D. Geiregat et al.,
 \PLB {\bf 247}, 131 (1990) and
    B {\bf 259}, 499 (1991);\\
CCFR Collaboration, C.G. Arroyo et al., \PRL {\bf 72}, 3452 (1994);
                    {\em Eur.~Phys.~J.} C {\bf 1}, 509 (1998) 
\bibitem{nutev}
NuTeV Collaboration, T. Bolton, talk at ICHEP98 (PS1), these proceedings
\bibitem{zfitter} D. Bardin et al., hep-ph/9412201
\bibitem{topaz0} G. Montagna, O. Nicrosini, F. Piccinini, 
                G. Passarino,
                hep-ph/9804211
\bibitem{bardinpassarino} D. Bardin, G. Passarino, hep-ph/9803425
\bibitem{langacker} J. Erler, P. Langacker, hep-ph/9809352; \\
                    hep-ph/9801422
\bibitem{higgsfits}
                K. Ha\-gi\-wa\-ra, D. Haidt, S. Ma\-tsu\-mo\-to,
                {\em Eur.\ Phys.\ J.\ } C {\bf 2}, 95 (1998); \\
                J. Ellis, G.L. Fogli, E. Lisi, \PLB {\bf 389}, 321 (1996);
                   \ZPC {\bf 69}, 627 (1996); \\
                   G. Passarino, {\em Acta Phys.~Polon.\ } {\bf 28}, 635 
                                  (1997); \\
                  S. Dittmaier, D. Schildknecht, \PLB {\bf 391}, 420 (1997);\\
                  S. Dittmaier, D. Schildknecht, G. Weiglein,
                  \PLB {\bf 386}, 247 (1996); \\
                  P. Chankowski, S. Pokorski, 
                  {\em Acta Phys.~Polon.\ } {\bf 27}, 1719 (1996)
\bibitem{deboer} W. de Boer, A. Dabelstein, W. Hollik, W. M\"osle,
                 U. Schwickerath, 
                 \ZPC {\bf 75}, 627 (1997)
\bibitem{kunszt}  Z. Kunszt et al., in
          {\em Physics at LEP 2}, CERN 96-01, Vol.\ 1, p.\ 141,  
           eds.\ G. Altarelli,
          T. Sj\"ostrand, F. Zwirner 
\bibitem{ww} W. Beenakker et al., in
          {\em Physics at LEP 2}, CERN 96-01, Vol.\ 1, p.\ 79,  
           eds.\ G. Altarelli,
          T. Sj\"ostrand, F. Zwirner
\bibitem{jadachetal} S. Jadach et al., \PLB {\bf 417}, 326 (1998); 
           S. Jadach et al, contributed paper to ICHEP98,
           abstract 823 (PS1)
\bibitem{argyres} E.N. Argyres et al., \PLB {\bf 358}, 339 (1995);\\
         U. Baur, D. Zeppenfeld, \PRL {\bf 75}, 1002 (1995);\\
         C. Papadopoulos, \PLB {\bf 352}, 144 (1995)
\bibitem{gabriel} G. Lopez Castro, J.L. Lucio M., J. Pestieau,
                {\em Int.\ J.~Mod.\ Phys.}~A {\bf 11}, 563 (1996); \\
         M. Beuthe, R. Gonzalez Felipe, G. Lopez Castro, J. Pestieau,
                \NPB {\bf 498}, 55 (1997)
\bibitem{fermionscheme} W. Beenakker at al., 
              \NPB {\bf 500}, 255 (1997)
\bibitem{beenakkerdenner} W. Beenakker, A. Denner,
          {\em Acta Phys.\ Polon.\ } B {\bf 29}, 2821  (1998)
\bibitem{nonfactor1} W. Beenakker, A.P. Chapovsky, F.A. Berends,
              \NPB {\bf 508}, 17 (1997);
              \PLB {\bf 411}, 203 (1997)
\bibitem{nonfactor2} A. Denner, S. Dittmaier, S. Roth,
              \NPB {\bf 519}, 39 (1998);
              \PLB {\bf 429}, 145 (1998). \\
 The original results of the earlier calculation \\
             K. Melnikov, O. Yakovlev, \NPB {\bf 471}, 90 (1996) \\
  does not agree with the other  two more recent results. As known
  from the authors, their corrected result is now also in agreement
  (O. Yakovlev, private communication; and Erratum, to appear) 
\bibitem{yakovlev}
          V.S. Fadin, V.A. Khoze, A.D. Martin, \PRD {\bf 49}, 2247 (1994); \\
          K. Melnikov, O. Yakovlev, \PLB {\bf 324}, 217 (1994)   
\bibitem{baur} U. Baur, talk at ICHEP98 (PS1), these proceedings; \\
              U. Baur, S. Keller, D. Wackeroth, hep-ph/9807417 
\bibitem{wackeroth} D. Wackeroth, W. Hollik, 
                   \PRD {\bf 55}, 6788 (1997)
\bibitem{passera} M. Passera, A. Sirlin, 
                {\em Acta Phys.\ Polon. } B {\bf 29} 2901 (1998) 
\bibitem{treille} D. Treille, plenary talk at ICHEP98, 
                  these proceedings 
\bibitem{lindner} L. Maiani, G. Parisi, R. Petronzio,
                 \NPB {\bf 136}, 115 (1979); \\
                  N. Cabibbo, L. Maiani, G. Parisi, R. Petronzio,
                   \NPB {\bf 158}, 259 (1979); \\ 
                 R. Dashen, H. Neuberger, \PRL {\bf 50}, 1897 (1983); \\
                 D.J.E. Callaway, \NPB {\bf 233}, 189; \\
                 M.A. Beg, C. Panagiotakopoulos, A. Sirlin,
                 \PRL {\bf 52}, 883 (1984); \\
                 M. Lindner, \ZPC {\bf 31}, 295 (1986) 
\bibitem{higgsbounds}
         M. Lindner, M. Sher, H. Zaglauer, 
                 \PLB {\bf 228}, 139 (1989); \\
        G. Altarelli, G. Isidori, \PLB {\bf 337}, 141 (1994); \\
        J.A. Casas, J.R. Espinosa, M. Quiros,
        \PLB {\bf 342}, 171 (1995) and 
           B {\bf 382}, 374 (1996); 
\bibitem{hambye} T. Hambye, K. Riesselmann, 
            \PRD {\bf 55}, 7255 (1997)
\bibitem{lattice} Kuti et al., \PRL {\bf 61}, 678 (1988);\\
        P. Hasenfratz et al., \NPB {\bf 317}, 81 (1989);\\
        M. L\"uscher, P. Weisz, \NPB {\bf 318}, 705 (1989); \\
        M. G\"ockeler, H. Kastrup, T. Neuhaus, F. Zimmermann,
        \NPB {\bf 404}, 517 (1993)
\bibitem{decaylattice}   
         M. G\"ockeler, H. Kastrup, J. Westphalen, F. Zimmermann,
          \NPB {\bf 425}, 413 (1994)
\bibitem{ghinculov}  A. Ghinculov, \NPB {\bf 455}, 21 (1995); \\
                   A. Frink, B. Kniehl, K. Riesselmann, 
                  \PRD {\bf 54}, 4548 (1996)
\bibitem{binoth} T. Binoth, A. Ghinculov, J.J. van der Bij,
             \PRD {\bf 57}, 1487 (1998); 
             \PLB {\bf 417}, 343 (1998)
\bibitem{Hff} L. Durand, B.A. Kniehl, K. Riesselmann,
             \PRL {\bf 72}, 2534 (1994); 
              E: {\em ibidem} B {\bf 74}, 1699 (1995); \\
          A. Ghinculov, \PLB {\bf 337}, 137 (1994); 
          E: {\em ibidem} B  {\bf 346}, 426 (1995); \\
          V. Borodulin, G. Jikia, \PLB {\bf 391}, 434 (1997) 
\bibitem{riess} K. Riesselmann, hep-ph/9711456
\bibitem{bsgammaQCD} K. Adel and Y.P. Yao, \PRD {\bf 49}, 4945
(1994); \\
C. Greub and T. Hurth, \PRD {\bf 56}, 2934 (1997);\\
A.J. Buras, A. Kwiatkowski and N. Pott, 
\PLB {\bf 414}, 157 (1997); \NPB {\bf 517}, 353 (1998);\\
A. Ali and C. Greub, Phys.\ Lett.\ B {\bf 361}, 146
(1995);\\
C. Greub, T. Hurth and D. Wyler, 
        \PLB {\bf 380}, 385 (1996); 
         \PRD {\bf 54}, 3350 (1996); \\
K. Chetyrkin, M. Misiak and M. M\"unz, 
\PLB {\bf 400}, 206 (1997); E: {\bf 425}, 414 (1998)
\bibitem{bsgammaEW}
A. Czarnecki and W.J. Marciano, \PRL {\bf 81}, 277  (1998);\\
 A.L. Kagan and M. Neubert, hep-ph/9805303
\bibitem{neubert} M. Neubert, talk at ICHEP98 (PS7), these proceedings;
                hep-ph/9809377
\bibitem {CLEO} T. Skwarnicki (CLEO Collaboration), 
talk at ICHEP98 (PS7), these proceedings
\bibitem {ALEPH} R. Barate et al.\ (ALEPH Collaboration),
          \PLB {\bf 429},  169 (1998)
\bibitem{newphys}
   J. Hewett, M. Neubert, 
   talks at ICHEP98  (PS16), these proceedings
\bibitem{brookhaven} V.W. Hughes, in: {\em Frontiers of High Energy
          Spin Physics}, Universal Academy Press, Tokyo 1992,
          p.~717, ed.\ T. Hasegawa  
\bibitem{ew1} K. Fujikawa, B.W. Lee, A.I. Sanda, \PRD {\bf 6},
          2923 (1972);\\
          R. Jackiw, S. Weinberg, \PRD {\bf 5}, 2473 (1972);\\
          G. Altarelli, N. Cabibbo, L. Maiani, \PLB {\bf 40}, 415
          (1972); \\
          I. Bars, M. Yoshimura, \PRD {\bf 6}, 374 (1972);\\
          W. Bardeen, R. Gastmans, B. Lautrup, \NPB {\bf 46}, 315
          (1972)
\bibitem{ew22} S. Peris, M. Perrottet, E. de Rafael,
         \PLB {\bf 355}, 523 (1995)
\bibitem{ew23} A. Czarnecki, B. Krause, W. Marciano,
         \PRD {\bf 52}, 2619 (1995); \PRL {\bf 76}, 3267 (1996)
\bibitem{degrassigiudice}
         G. Degrassi,  G.F. Giudice, hep-ph/9803384
\bibitem{dub} K. Adel, F.J. Yndurain, hep-ph/9509378; \\
          D.H. Brown,W.A. Worstell, \PRD {\bf 54}, 3237 (1996)
\bibitem{sanda} M. Hayakawa, T. Kinoshita, A.I. Sanda,
         \PRL {\bf 75}, 790 (1995);
         \PRD {\bf 54}, 3137 (1996)
\bibitem{hayakawa} M. Hayakawa, T. Kinoshita, 
          \PRD {\bf 57}, 7267 (1997)     
\bibitem{bijnens} J. Bijnens, E. Pallante, J. Prades,  \NPB {\bf 474},
                  379 (1996);
                  \PRL {\bf 75}, 1447 (1995); 
                  E: {\em ibidem} {\bf 75}, 3781 (1995)
\bibitem{had3} B. Krause, \PLB {\bf 390}, 392 (1997); \\
          T. Kinoshita, B. Nizic, Y. Okamoto,
          \PRD {\bf 31}, 2108 (1985); \\
          T. Kinoshita, W. Marciano, in: {\em Quantum Electrodynamics},
          ed.\ T. Kinoshita, World Scientific 1990 (p.\ 419)
\bibitem{qed} T. Kinoshita, \PRL {\bf 75}, 4728 (1995);\\
          S. Laporta, E. Remiddi, \PLB {\bf 379}, 283 (1996)
\bibitem{neutron} B. Krause, A. Czarnecki, \PRL {\bf 78}, 4339 (1997)
\bibitem{superka} 
 Y. Fukuda et al., hep-ex/9805006; \\ hep-ex/9805021; hep-ex/987003; \\
  M. Takita, plenary talk at ICHEP98, these proceedings
\bibitem{hmass} M. Carena, J. Espinosa, M. Quiros, C. Wagner,
                \PLB {\bf 355}, 209 (1995; \\
                 M. Carena, M. Quiros, C. Wagner,
                \NPB {\bf 461}, 407 (1996); \\
                H. Haber, R. Hempfling, A. Hoang,
                \ZPC {\bf 75}, 539 (1997); \\
                S. Heinemeyer, W. Hollik, G. Weiglein,
                \PRD {\bf 58},  091701 (1998); 
                hep-ph/9807423 
                ({\em Phys.\ Lett. }B, to appear)
\bibitem{higgs} A. Denner, R. Guth, W. Hollik, J.H. K\"uhn,
                \ZPC {\bf 51}, 695 (1991);\\
                J. Rosiek, \PLB {\bf 252}, 135 (1990;\\
                M. Boulware, D. Finnell, \PRD {\bf 44}, 2054 (1991)
\bibitem{susy1}  G. Altarelli, R. Barbieri, F. Caravaglios,
                \PLB {\bf 314}, 357 (1993); \\
                C.S. Lee, B.Q. Hu, J.H. Yang, Z.Y. Fang,
                {\em J. Phys.\ } G {\bf 19}, 13 (1993); \\
                Q. Hu, J.M. Yang, C.S. Li, {\em Commun.\ Theor.\ Phys.\ }
                {\bf 20}, 213 (1993); \\
                J.D. Wells, C. Kolda, G.L. Kane, \PLB {\bf 338}, 219 (1994); \\
                G.L. Kane, R.G. Stuart, J.D. Wells,
                \PLB {\bf 354}, 350 (1995);\\
                M. Drees et al., \PRD {\bf 54}, 5598 (1996)
\bibitem{susydelr} P. Chankowski, A. Dabelstein, W. Hollik, W.~M\"osle,
                   S. Pokorski, J. Rosiek, \NPB {\bf 417}, 101 (1994); \\
                   D. Garcia,  J. Sol\`a, {\em Mod.\ Phys.\
                   Lett.\ } A {\bf 9}, 211 (1994)
\bibitem{susy3} D. Garcia, R. Jim\'enez, J. Sol\`a,
                \PLB {\bf 347}, 309 and 321 (1995);\\
                D. Garcia, J. Sol\`a, \PLB {\bf 357}, 349 (1995);\\
                A. Dabelstein, W. Hollik, W. M\"osle, in {\em
                Perspectives for  Electroweak
                Interactions in $\epm$ Collisions},
                Ringberg Castle 1995,
                ed.\ B.A. Kniehl, World Scientific 1995 (p.\ 345);\\
                P. Chankowski, S. Pokorski, \NPB {\bf 475}, 3 (1996); \\
                J. Bagger, K. Matchev, D. Pierce, R. Zhang,
                \NPB {\bf 491}, 3 (1997)
\bibitem{drosusy} A. Djouadi, P. Gambino, S. Heinemeyer, W. Hollik,
                  C. J\"unger, G. Weiglein,
                  \PRL {\bf 78}, 3626 (1997); \PRD {\bf 57}, 4179 (1998) 
\bibitem{pierce} J. Erler, D. Pierce, \NPB {\bf 526}, 53 (1998)

\end{thebibliography}
